\documentclass[twocolumn]{aastex631}

\usepackage[whole]{bxcjkjatype}
\usepackage{graphicx}
\usepackage{xcolor}
\usepackage{amsmath, amssymb}
\usepackage[normalem]{ulem} 
\usepackage{multirow}
\usepackage{hyperref}
\usepackage[colorinlistoftodos,backgroundcolor=yellow,textsize=tiny,textwidth=1.25cm]{todonotes}
\usepackage{comment}
\usepackage{bold-extra}

\newcommand{\rin}{{r_\mathrm{in}}}
\newcommand{\rout}{{r_\mathrm{out}}}

\newcommand{\bb}{{\mathbf{b}}}
\newcommand{\vv}{{\mathbf{v}}}

\newcommand{\Bv}{{\mathbf{B}}}

\newcommand{\lfrac}[2]{{{#1}/{#2}}}

\newcommand{\cm}{{\mathrm{\,cm}}}
\newcommand{\cmNSP}{{\mathrm{cm}}}
\newcommand{\second}{{\mathrm{\,s}}}

\newcommand{\km}{{\mathrm{\,km}}}

\newcommand{\kms}{{\km\second^{-1}}}

\newcommand{\diffu}{{\cm^2\second^{-1}}}

\newcommand{\au}{{\mathrm{\,au}}}

\newcommand{\yr}{{\mathrm{\,yr}}}

\newcommand{\gm}{{\mathrm{\,g}}}

\newcommand{\gram}{{\gm}}

\newcommand{\massden}{{\gram\cm^{-3}}}

\newcommand{\numden}{{\cm^{-3}}}

\newcommand{\rsun}{{R_\sun}}

\newcommand{\K}{{\mathrm{\,K}}}

\begin{document}

\title{Resistive Collapse of 2D Non-rotating Magnetized Isothermal Toroids: Formation of Pseudodisks}

\author[0009-0007-5437-2722]{Ya-Chi Wang (王亞琪)}
\affiliation{Institute of Astronomy and Astrophysics, Academia Sinica,  Taipei 106216, Taiwan}

\author[0000-0001-8385-9838]{Hsien Shang (尚賢)}
\affiliation{Institute of Astronomy and Astrophysics, Academia Sinica,  Taipei 106216, Taiwan}

\author[0000-0001-5557-5387]{Ruben Krasnopolsky}
\affiliation{Institute of Astronomy and Astrophysics, Academia Sinica, Taipei 106216, Taiwan}

\correspondingauthor{Hsien Shang}
\email{shang@asiaa.sinica.edu.tw}

\begin{abstract}
The collapse of singular magnetized toroids \citep{li1996b} is a natural representation of an early phase in star formation, bridging the prestellar and protostellar phases of the collapse of molecular cloud cores. We revisit the collapse study of \citet{allen2003}, now with explicit nonideal MHD (Ohmic diffusivity $\eta$) and higher resolution using a code able to cover a broader range of the magnetization parameter $H_0$.
Galli-Shu equatorial pseudodisks form for all values of $H_0$ and $\eta$, and the asymptotic central mass growth rate is in the scale $\Dot{M}_*\sim(a^3/G)(1+H_0)$, where $a$ is the isothermal sound speed, consistent with previous results and predictions.
The explicit Ohmic diffusivity makes the field line structure less radial than in previous work, connecting the pseudodisk more effectively to its surroundings. Matter can fall efficiently onto the pseudodisk surfaces, forming oblique shocks, where shock heating and large density gradients raise the possibility of rich astrochemistry.
Pseudodisk size and structure are influenced by magnetic diffusivity.
Force and velocity ratios were computed to explore the magnetic support within the pseudodisk and its induced slowdown in infall velocity. Magnetic diffusivity was measured to control the strength of these effects and their location within the pseudodisk.
The dependence of the field line configurations, pseudodisk structure, and velocity ratios on magnetic diffusivity has observable consequences for collapsing envelopes.
\end{abstract}


\section{Introduction} \label{sec:intro}

Stars are born in star-forming regions within molecular clouds.  This process involves several stages \citep{shu1987}.
Clumps within the cloud lose their magnetic and turbulent support through ambipolar diffusion, resulting in the formation of the molecular cloud cores, which evolve centrally condensed states (\citealt{shu1977}, \citealt{bodenheimer1968}, \citealt{li1996b}, \citealt{lizano1989}).
These centrally condensed core states are unstable and collapse under their own gravity. A central protostellar object forms \citep{das2025}. This object is on the order of several $\rsun$, sufficiently small to be modeled in this work as a point mass, surrounded by the evolving remains of the centrally condensed states. 
In this stage, the collapse becomes fully dynamical, progressing from the inside-out. Material falls in, and accretes through pseudodisks and rotationally supported disks into the central protostar \citep{li2014,vaisala2023}.
Jets with wide-angle winds \citep{shu1994_XWI,blandford1982} and outflows are generated \citep[e.g.,][]{basu2024} and interact with their environments \citep{shang2020}.

In this work we focus on a model of the transitional stage between the condensed prestellar stage, and the beginning of point-mass formation.
The condensed core collapses, and a central protostellar mass starts to grow. A wave of dynamical infall propagates from the inside out. Within this expanding wave of collapse, matter falls in. In a hydrodynamic model \citep{shu1977}, this infall could be spherically symmetric, the central mass would grow at a rate $\Dot{M}_*\sim a^3/G$, where $a$ is the isothermal sound speed, and the inner parts of the collapse would be near free fall.
A magnetized model will break the spherical symmetry.
As ionized gas falls in, it drags poloidal magnetic field lines inwards. The dragged field lines bend, and create intense magnetic tension near the equator.
This magnetic tension slows down the gas infall in the equatorial areas, causing density to increase and mass to accumulate and form a pseudodisk. This flattened disk-like structure is partially supported by magnetic tension against gravity \citep{galli1993_i,galli1993_ii}.
In an ideal MHD model of the process, field lines and mass infall are perfectly tied to each other: the dragging of field lines would be very intense and the inner magnetic configuration would become radial \citep{galli2006}. A more realistic nonideal MHD model allows partial coupling between mass and magnetic field. Magnetic diffusion allows matter to fall in across field lines, resulting in a magnetic configuration without a central singular point \citep{shu2006}.
Infall velocities within the pseudodisk are slower than in the hydrodynamic model, by a factor related to the partial magnetic support \citet{shu1997_isopedic}.

Numerical simulations \citep[e.g.,][]{allen2003,mellon2009,vaisala2023} confirm this picture of the formation of pseudodisks as an integral part of molecular cloud core collapse.
Observational evidence has firmly established their existence.
The pioneer observations of pseudodisks include \citet{ohashi1991,
hayashi1993,ohashi1995,ohashi1996_L1551,ohashi1996_nobeyama,ohashi1997}.
More recent observations include the collapsing inflow and magnetic field structures of B335 \citep[e.g.][]{saito1999, yen2015, maury2018, yen2020, evans2023}, HH 211 \citep[e.g.][]{lee2014_hh211,lee2019,yen2023}, HH 212 \citep[e.g.][]{lee2014_hh212}, and L1157-mm \citep[e.g.][]{looney2007,chiang2010,stephens2013}, which can be well described by the pseudodisk concept of structures partially magnetically supported and partially infalling during the process of molecular cloud core collapse.

\citet{liPP6} and \citet{TsukamotoPP7} and references therein give an overall picture of more recent theoretical works on aspects of pseudodisk and disk formation.
In the presence of rotation, magnetic braking flows were studied in \citet{allen2003_ii}, and following up on that, \citet{mellon2008,mellon2009} found that magnetic braking can make disk formation difficult, especially in axisymmetric ideal MHD, but also in certain nonideal MHD configurations.
This motivated study of disk formation simulations with nonideal MHD, regularly forming pseudodisks as well, e.g., in \citet{krasnopolsky2010,krasnopolsky2011}, \citet{li2011}, and \citet{zhao2016,zhao2020,zhao2021}. Magnetic misalignment and turbulence in non-axisymmetric pseudodisk and disk formation are treated in, e.g., \citet{li2013,li2014}, \citet{vaisala2019}, and \citet{wang2022}. \citet{zhao2011} and \citet{krasnopolsky2012} study magnetic redistributions that take place during these collapse processes. Details of the nonideal MHD physics are explored in \citet{zhao2018}. The nonideal MHD simulations of accreting disks in \citet{suriano2017,suriano2018,suriano2019} show structure with rings and gaps. \citet{vaisala2023} have performed 3D simulations of pseudodisk formation surrounding a central mass in the presence of fluid viscosity and a density-dependent Ohmic resistivity. More recently, the protostellar masses formed during collapse in \citet{das2025} are regularly surrounded by flattened magnetized structures akin to those in \citet{galli2006}, describable as thin portions of pseudodisks.

This study aims to revisit the classical study of toroid collapse of \citet{allen2003}.
We explicitly include the effects of Ohmic resistivity. Using modern hardware and updated simulation software, we explore a larger parameter space at a higher resolution.
We can now compute the collapse of toroids in a broader range of the magnetization support parameter $H_0$, and examine the explicit dependence of the collapse on the level of magnetic diffusivity $\eta$.
The simulations use the singular isothermal toroids \citep{li1996b} as initial conditions, and evolve them using nonideal magnetohydrodynamic (MHD) with explicit Ohmic resistivity.
In this new study, we confirm classical results and find new features.
The asymptotic values of central mass growth rate $\Dot{M}$ confirm predictions and results \citep{li1996b,allen2003}, but the timescale of asymptotic approach acquires a dependence on $\eta$ (\S\ref{sec:diffusion_time}).
The absence of rotation in this study limits its scope, because magnetic braking and magnetocentrifugal flows are not generated. We will next follow up \citet{allen2003_ii} to extend the scope of this study.

The observable infall within the pseudodisk is slowed down by magnetic tension. We find that the microphysical parameter $\eta$ controls the ratio of magnetic to gravity forces, and therefore the infall slow down ratio $v_r/v_\text{ff}$. Within the infall region $r<at$, the shape of the field lines and their ability to exert magnetic tension is controlled not only by their initial magnetization $H_0$ but also by the diffusivity $\eta$.

We notice shocks at the pseudodisk surfaces, which are enabled by the diffusivity enhancing the ability of matter to directly fall onto the pseudodisk along non-radial field lines. The presence of shocks is not limited to rotationally supported disks. This has potentially observable consequences in astrophysics and astrochemistry. Velocity and density contours serve to visualize the shocks and the shape and growth of the pseudodisk.

This work is a foundation for further exploration of nonideal MHD effects with the isothermal toroids at higher resolutions, filling the gap between the stationary analytic models and the actual evolved protostellar envelopes influenced by the gravity from the growing central protostar and the self-gravitation of the evolving structures. These simulated results mimic actual observational data, allowing for comparison and synthetic observations \citep[e.g.,][]{lee2019, shang_PII,ai2024, liu2025}. 

This paper is organized as follows. In \S\ref{sec:theory} we review some theory results used in this work.
We next introduce the methodology in \S\ref{sec:numerical_methods}, and
report our numerical results in \S\ref{sec:results}. Further discussion and analysis are presented in \S\ref{sec:discussion}, followed by a summary (\S\ref{sec:summary}).

\section{Theoretical Background} \label{sec:theory}
We utilize the resistive, nonideal magnetohydrodynamics equations 
\begin{equation}\frac{\partial\rho}{\partial t}+\nabla \cdot (\rho\vv)=0\end{equation}
\begin{equation}\frac{\partial\vv}{\partial t}+(\vv\cdot \nabla )\vv=-\frac{\nabla p }{\rho }+\frac{(\nabla \times\Bv)\times\Bv}{4\pi\rho}-\nabla \Phi_g\end{equation}
\begin{equation}\frac{\partial\Bv}{\partial t}=\nabla \times (\vv\times\Bv-\eta \nabla \times\Bv)\end{equation}
where $\rho$ is the density, $\vv$ is the velocity, $\Bv$ is the magnetic field, $\Phi_{g}$ is the gravitational potential, and $\eta$ is the Ohmic diffusivity.
We study a phase in star formation in which cooling processes are expected to be efficient, and therefore write the pressure $p=a^2\rho$ using the isothermal equation of state with sound speed $a$.
Spherical coordinates $(r, \theta, \phi)$ and cylindrical coordinates $(z, \varpi, \phi)$ are used throughout this work. We assume axisymmetry in this study.
In addition to the field quantities, the system also has a central point mass $M_*(t)$ located at $r=0$.
The gravitational potential is the sum
\begin{equation}
\Phi_g=\Phi_{g, \text{central}}+\Phi_{g, \text{self}}
\end{equation}
of the central potential $\Phi_{g, \text{central}}=-GM_*/r$, and the self-gravity potential, solution of the Poisson equation 
\begin{equation}
    \nabla ^2 \Phi_{g, \text{self}} =4\pi G\rho \label{eq:Poisson}
\end{equation}
where $G$ is Newton's gravity constant.

\subsection{Singular Isothermal Spheres} \label{sec:sphere_collapse}
During the collapse of isothermal cloud cores, a few phases naturally take place \citep[][Chapter 18]{shu1992}, bridging the late prestellar phase and the early protostellar phase of the core collapse.

The prestellar phase evolves in a quasistatic manner, in which the cloud core goes through a series of partial equilibria, supported against self-gravity by gas pressure and magnetization. This phase features a slow collapse (Figure 1 of \citealt{shu1977}). Starting, e.g., from a uniform density profile, the inner parts become progressively denser, while the outer parts acquire a density profile $\propto r^{-2}$. This power law feature is regularly observed to develop in the outer portions of calculations and numerical simulations \citep{bodenheimer1968}.
Near the end point of the prestellar phase, the whole structure has the $r^{-2}$ power law. The natural outcome of this quasistatic phase is the power law density distribution $\rho (r)=a^2/2\pi Gr^2$, the singular isothermal sphere in hydrostatic equilibrium.

The collapse of the singular isothermal spheres proceeds in a self-similar manner.
It starts at time $t=0+$ by the formation and growth of a central point mass, which represents the beginning of the formation of a star. The self-similar scalings reappear in the collapse of magnetized toroids, and so we present them here in the form of the dimensionless variables
\begin{align}
x&=r/at\label{eq:SISx}\\
\rho (r, t)&=\alpha (x)/4\pi Gt^2\\
M(r, t)&=t (a^3/G) m(x)\\
u(r, t)&=av(x),\label{eq:SISv}
\end{align}
where $M(r, t)$ is the enclosed mass within radius $r$ at time $t$, and $u$ is the dimensional radial velocity (Equations 7 and 8 in \citealt{shu1977}). The three functions $\alpha$, $m$, and $v$ are respectively the density, enclosed mass, and radial velocity in dimensionless units, solutions of the system of equations (10--12) in \citet{shu1977}.
The dimensionless enclosed mass at zero radius is $m_0\equiv m(x\!=\!0)$, representing the central point mass as
$M_*(t)=M(r\!=\!0,t)=t (a^3/G)m_0$.
The initial conditions at $t=0^+$ are the asymptotic solutions for large $x$ given by 
 $\alpha\rightarrow A/x^2$, $v=-(A-2)/x$, where the hydrostatic sphere able to collapse has $A=2^+$.

The unique solution for $A=2^+$ represents the collapse of a critical singular isothermal sphere.
This solution features a self-similar expansion wave with its front located at $x=1$, that is, $r=a t$, propagating at the sound speed. At large radii ($x>1$, $r>a t$) outside the expansion wave, the flow is still unperturbed and it still has the density profile $\propto r^{-2}$.
At small radii ($x\ll1$, $r\ll a t$) within the innermost parts of the expansion wave, the enclosed mass $M(r,t)$ approaches its central value $M_*=t (a^3/G) m_0$, where $m_0=0.975$ for $A=2^+$.
In this region, $x\ll1$, the flow reaches free fall, and the dimensionless speed and density respectively asymptote to $v=-(2m_0/x)^{1/2}$ and $\alpha=(m_0/2x^3)^{1/2}$.
The dimensionless mass enclosed within the expansion wave, from $x=0$ to $x=1$, is $m(x\!=\!1)=2$. Therefore, the total mass within the inside-out collapse is $M(r\!=\!at,t)=2 t (a^3/G)$, with almost half ($0.49=m_0/2$) of this mass corresponding to $M_*$. Self-gravity is approximately equally as important as central gravity in the scale of the whole inside-out collapse region. In contrast, the central mass gravity dominates close enough to the center, and the flow reaches free fall.

\subsection{Singular Isothermal Toroids} \label{sec:toroids}

The magnetized singular isothermal toroid solutions found in \citet{li1996b} are a natural extension of the quasistatic phase of the singular isothermal spheres. In the universally expected presence of magnetization, the quasistatic approach leads to prestellar structures supported against self-gravity by combining magnetic and thermal forces. The magnetic field is not isotropic. Hence, these structures are not spheres but axisymmetric toroids.

Extensive literature has shown that prestellar cores evolve quasi-statically. Density profiles evolve to $\rho\propto r^{-2}$ \citep[e.g.][]{lizano1989}, and magnetic field intensities evolve to $B\propto r^{-1}$ \citep[e.g.][]{fiedler1993,basu1994}.  These power-laws are applied  to calculate the density and magnetic flux of the toroids in the form presented in \citet{li1996b}:
\begin{align}
\rho(r, \theta )&=\frac{a^2}{2\pi Gr^2}R(\theta )\label{eq:toroid_rho}\\
\Phi(r,\theta )&=\frac{4\pi a^3 r}{G^{1/2}}\phi (\theta )\label{eq:toroid_flux}.
\end{align}
The functions $R(\theta)$ and $\phi(\theta)$ are obtained by
balancing self-gravity, isothermal pressure, and magnetic forces in quasi-equilibrium, expressed as a set of coupled ODEs for the functions $\phi(\theta)$ and $R(\theta)$ (Equations 12 and 13 in \citealt{li1996b}), solved in the domain $0\leq\theta\leq\pi/2$. The equatorial boundary conditions are that density contours and field lines are symmetric and smooth at the equator. The axial boundary conditions are that the magnetic flux at the axis is zero, and that there is no axial line mass.
The magnetic field is poloidal and obtained from the magnetic flux function.
Integrating $\rho$ for the volume up to a radius $r$, with these boundary conditions,
and using that for this prestellar calculation, the central mass is set to $M(r\!=\!0)=0$, the mass enclosed within a radius $r$ is
\begin{equation}\label{eq:toroid_enclosed}
    M(r)= r (2a^2/G) (1 + H_0),
\end{equation}
where $H_0>0$ is a parameter in the equations, a function of the mass to flux ratio $\lambda$, and going to zero in the nonmagnetic limit.
Equation (\ref{eq:toroid_enclosed}) shows that $H_0$ measures the fraction of the toroid mass supported by magnetization, which is added to the value $M(r)= r (2a^2/G)$ obtained for a hydrodynamic prestellar singular isothermal sphere, whose support is only gas pressure.
The parameter $H_0$ also controls the opening and flattening of the toroids.
In the asymptotic limit $\theta\rightarrow0$, the dimensionless density behaves as $R(\theta)\rightarrow\theta^n$, where $n\equiv 4 H_0$.
The limiting case $n=0$ corresponds to the singular isothermal sphere with no magnetic flux, and $R(\theta)=1$.
In the opposite limit $H_0\gg1$, the solution becomes an equatorial thin disk in magnetostatic equilibrium, threaded by a poloidal magnetic field.
In the general case between these limits, the solutions are toroid-shaped, with lower densities near the axis $\theta=0\arcdeg$ and higher densities near the equator $\theta=90\arcdeg$. The toroids become more flat near the equator and more open near the axis as $n$ increases.

\citet{li1996b} also contains a dimensional estimate for the infall rate of these equilibrium toroids once they start collapsing in the protostellar phase. Then the central parts will be in a free fall collapse and feed the growing central mass at a rate $\Dot{M}_*=\mathcal{M}_0(1+H_0) a^3/G$ where $\mathcal{M}_0=0.975$ for small $H_0$ \citep{shu1977}, and $\mathcal{M}_0=1.05$ for the $H_0\gg1$ limit, based on the results of \citet{li1997}.

\subsection{Ideal MHD collapse and Numerical considerations} \label{sec:toroid_collapse}
\citet{allen2003} have started the investigations into the gravitational collapse of the magnetized toroids of \citet{li1996b} using versions of the Zeus code incorporating ideal MHD and self-gravity, but without explicitly considering magnetic diffusivity.
They examined the collapse of non-rotating toroids with $H_0=0.125$, $0.25$, and $0.5$.

As expected from \citet{galli1993_i,galli1993_ii}, the collapse of the toroids formed a flattened pseudodisk near the equator.
Within this structure, magnetic field lines are dragged by the infalling gas, creating strongly pinched magnetic configurations.
Magnetic tension partly supports the pseudodisk gas against gravity, locally decelerating the collapse, and enhancing mass density.
Magnetic forces exerted by the pinched field lines compress the pseudodisk and contribute to make it thinner. In the simulations reported in \citet{allen2003}, the denser parts of the pseudodisk are very thin and not fully resolved in the $z$ direction.

Matter from the low-density axial regions falls along field lines towards the equator and feeds the pseudodisk growth.
The collapse also forms a central point-mass $M_*$, whose dimensionless growth rate $m_0=\Dot{M}_* / (a^3/G)$ approaches asymptotic values within 2\% of $0.975(1+H_0) a^3/G$, consistent with small-$H_0$ predictions in \citet{li1996b}. However, \citet{allen2003} \S3.2 noted that the numerical coefficient preceding the $(1+H_0)$ factor tends to decrease with $H_0$ instead of increasing (an increase to $1.05$ is expected for $H_0\gg1$, \citealt{li1997}).

The simulations of \citet{allen2003} were very demanding.
In ideal MHD without resistivity, the exact analytical solution for collapse preserves the mass-to-flux ratio. Upon formation of a central point mass, magnetic flux concentrates at the origin, resulting in an ideal MHD configuration with extremely high central $B$-values and strong equatorial magnetic tension forces. In the numerical simulations, the extreme values were controlled by numerical effects and technique. Performed in cylindrical coordinates, their simulations reserved a central sink cell to represent central mass, which limited the magnetic field and introduced numerical diffusivity. As noted in their \S3.2, numerical diffusivity in their code mimicked some effects of magnetic diffusivity.
They describe a ``magnetic barrier'' oscillation that hindered achieving high values of the magnetization parameter $H_0$. This oscillation alternates between two field line topologies: a central configuration with field lines reaching the central sink cell, and a reconnected configuration where all field lines cross the equator. Our simulations in this work consistently exhibit the reconnected configuration, while Figure 4 of \citet{allen2003} approximates the central configuration. Starting from a central configuration, inevitable numerical (or physical) diffusivity allows field lines and magnetic flux to diffuse from the center into the pseudodisk's inner regions as reconnected field lines. This diffused field temporarily increases magnetic tension, briefly slowing accretion. Mass accumulates behind the reconnected field lines until gravity overcomes magnetic support, causing mass infall and dragging field lines back to the central configuration. Our simulations avoid these oscillations because, for the $\eta$-range and grid presented (\S\ref{sec:numerical_methods}), the reconnected configuration becomes stabilized. Mass flows across field lines due to magnetic diffusivity, falling towards the center, and does not pile up. Our physical magnetic diffusivity is large enough to dominate numerical diffusivity and to suppress the oscillations, enabling a broader exploration of $H_0$ and increased resolution. Oscillations can only appear and grow at physical diffusivity levels lower than those presented here.
A comparison of time scales illustrating this process is presented in \S \ref{sec:diffusion_time}.

Future work using nonideal MHD simulations with high order codes will achieve the best situation. Higher order codes such as Astaroth (a high order GPU code, \citealt{vaisala2023}) and Sadhana (a high resolution AMR code, \citealt{bandopadhyay2022}) can achieve much reduced numerical diffusivity. The lower numerical diffusivity will allow exploring an even wider range of values of physical diffusivity than in this work. Careful parameter choices will be set up to reduce the oscillations, without having to sacrifice any precision or accuracy.

\section{Numerical methods} \label{sec:numerical_methods}
The toroid collapse is computed using the ZeusTW code \citep{krasnopolsky2010}, equipped with nonideal MHD and self-gravity.
Simulations are performed in axisymmetric spherical coordinates, with a computational domain extending in $\theta$ from $0$ to $\pi=180\arcdeg$, and in radius $r$ from $\rin=4\au$ to $\rout=10^4\au$.
The isothermal sound speed $a=0.2\kms$ sets the units of density and magnetic flux according to Equations 
(\ref{eq:toroid_rho}) and (\ref{eq:toroid_flux}). The dimensionless functions $R(\theta)$ and $\phi(\theta)$ are given by the solution of the toroid ODE, which depends on the value of the parameter $n=4H_0$. The values of $n$ chosen for this set of simulations are $1$, $2$, and $4$. A tiny spherically symmetric density ($\rho_S=D^S/r^2$ with $D^S=2.25\times10^{11}\gram\cm$) is added in the initial setup to prevent the Alfv\'en speed near the axis from diverging \citep{shang2020}.

The outflow boundary conditions are applied at the radial boundaries, allowing matter to flow out of the grid, and reflection is applied at the two axial boundaries in the $\theta$ direction. 
The simulation, in addition to the fields defined within its computational domain, also has a central mass $M_*$, modeled numerically as a sink particle located at $r=0$. That mass $M_*$ grows with time as matter falls in through the computational boundary at $r=\rin$, and it has a (rather tiny) initial value equal to the mass (outside the computational domain) that would have been hypothetically present in the toroid for $r<\rin$, given by Equation (\ref{eq:toroid_enclosed}) as
\begin{equation}
    M_*(t=0)=\rin (2a^2/G) (1 + H_0),
\end{equation}
ranging from $8.99\times10^{29}\gram$ (for $n=1$) to $1.438\times10^{30}\gram$ ($n=4$).
For this setting, no additional initial central mass was necessary to initiate collapse.
The effects of explicit resistivity have been explored with three values of magnetic diffusivity, $\eta=10^{18}$, $10^{19}$, and $10^{20}\diffu$.
These values are based on the models of \citet{krasnopolsky2010} which show characteristic behavior (disk formation in their context) starting at $\eta=3\times10^{18}$--$3\times10^{19}\diffu$, and they can be compared with the fiducial values $\sim 10^{20}\diffu$ found for the magnetically diffusive behavior in \citet{shu2006}. These values are also compared in Section \ref{sec:diffusion_time} with estimates of $\eta_\text{AD}$ in \citet{TsukamotoPP7}, which are in the range $\gtrsim 2\times10^{18}\diffu$.

While simulations are performed in dimensional units, we report some of the results in dimensionless units, using the definitions as in 
Equation (3) of \citet{allen2003}
\begin{align}
x&={r}/{at}\\
\rho (r, \theta , t)&=\frac{\alpha (x, \theta , t)}{4\pi Gt^2}\\
\vv(r, \theta , t)&=a\bar{\vv}(x, \theta , t)\\
\Bv(r, \theta , t)&=\frac{a}{G^{1/2}t}\bb(x, \theta , t)
\end{align}
to define dimensionless density $\alpha$, dimensionless magnetic field $\bb$, and dimensionless velocity $\bar{\vv}=\vv/a$, dependent only on $x$ and $\theta$ in the self-similar limit. We reserve the letter $v$ for the dimensional velocity, a quantity called $u$ in \citet{allen2003} and in \citet{shu1977}.
The growth rate $\Dot{M}_*$ of the central mass is reported using the dimensionless mass infall rate defined in \citet{shu1977}, $m_0=\Dot{M}_*/(a^3/G)$, always of order unity in our results.

\section{Results} \label{sec:results}
We performed a set of nine simulations as described in \S\ref{sec:numerical_methods}.
The simulations started at the time $t=0$, which marks the transition between the quasi-static core evolution phase and the dynamic protostellar accretion phase.
We ran the simulations until or beyond the time at which the dimensionless functions showed signs of approaching a self-similar state.
The simulations yield comparable qualitative outcomes across all cases. Each scenario displays a smooth evolutionary process, marked by the inside-out collapse that initiates at the center and expands outward (as in \citealt{shu1977}), and by the formation of the pseudodisk \citep{allen2003}. All cases tend to develop toward a self-similar configuration. Their basic behaviors will be covered in this section and discussed in the next.

\begin{figure*}
    \includegraphics[width=\linewidth/2]{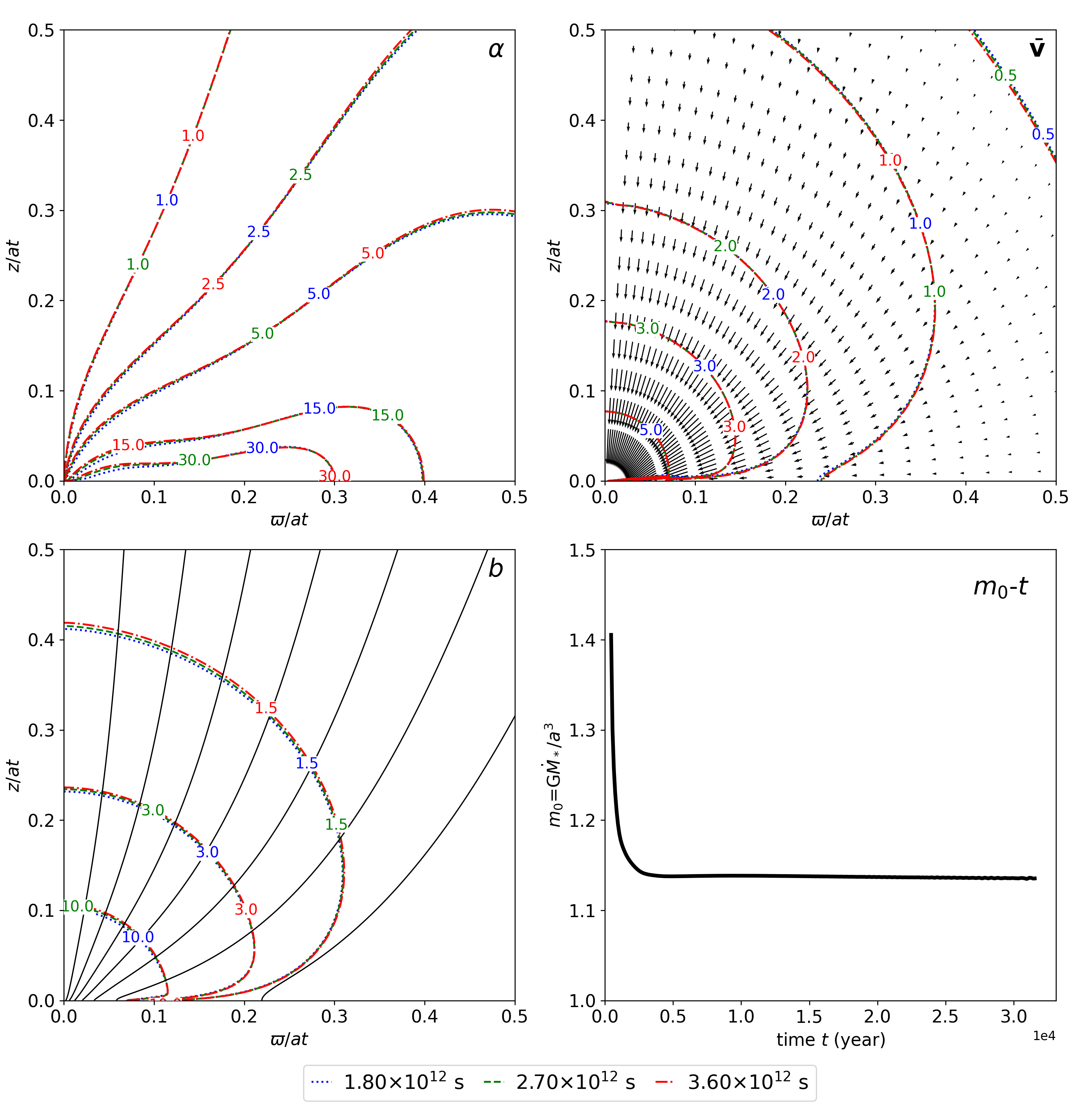}
    \includegraphics[width=\linewidth/2]{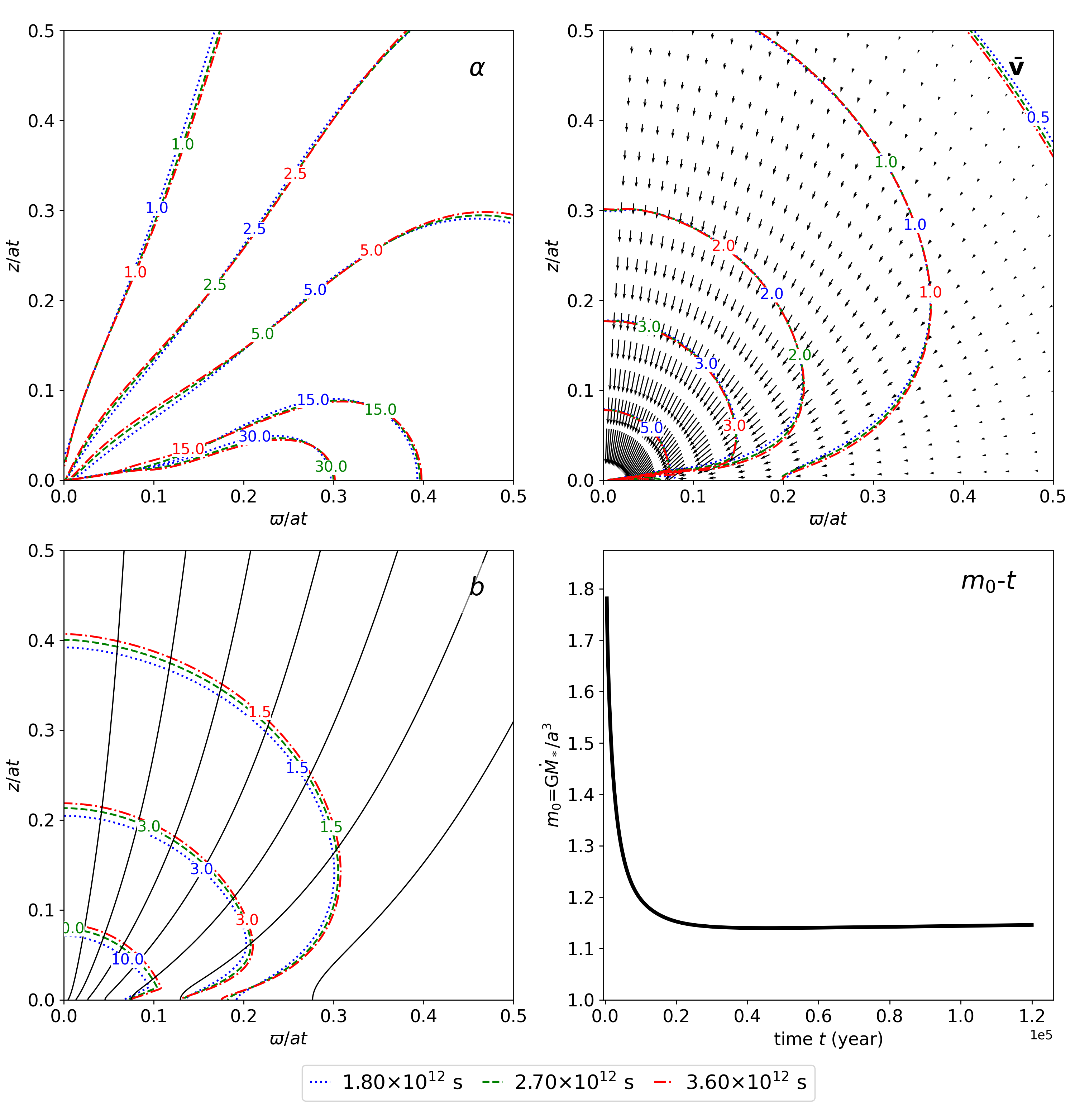}
    \includegraphics[width=\linewidth/2]{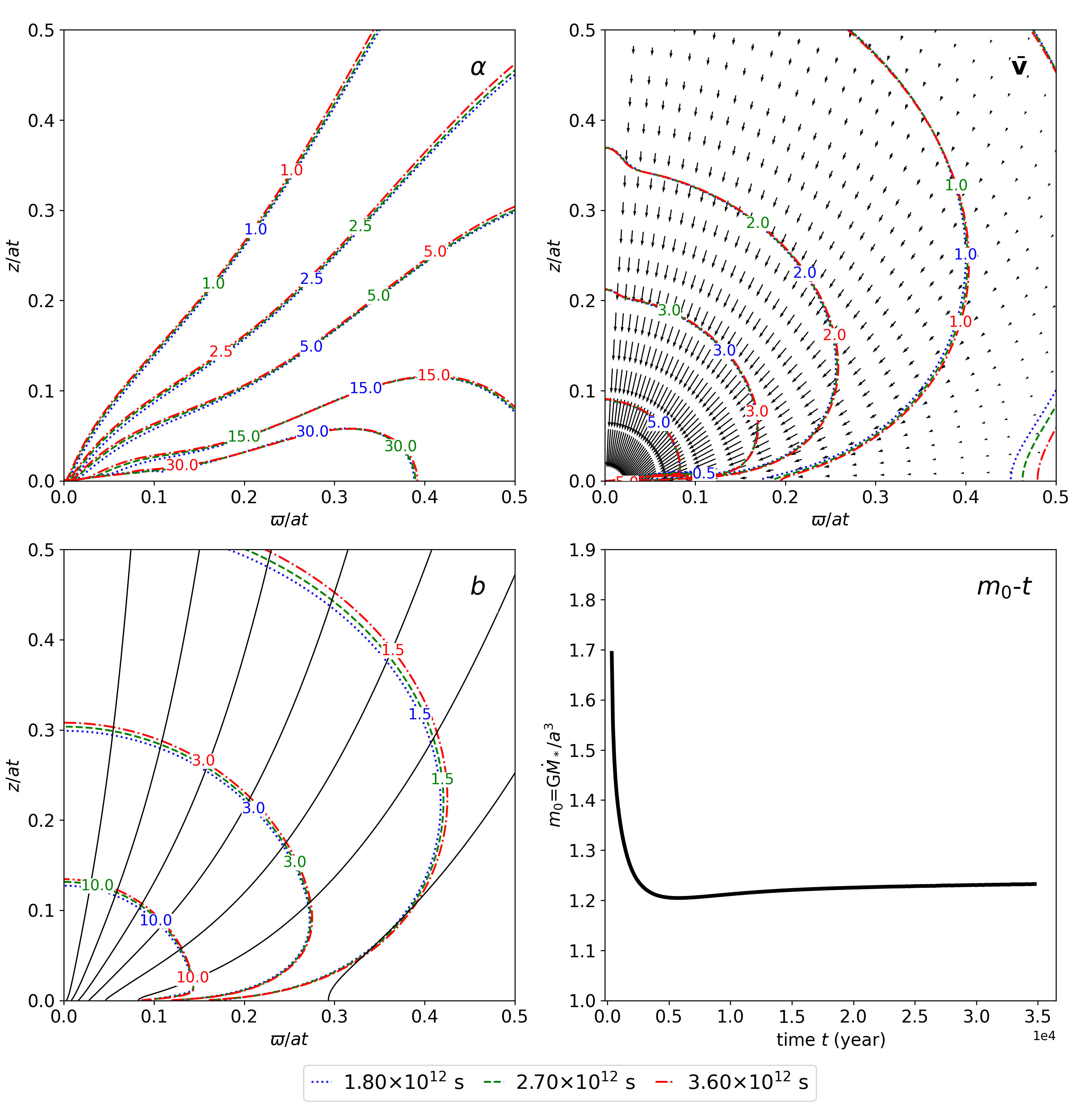}
    \includegraphics[width=\linewidth/2]{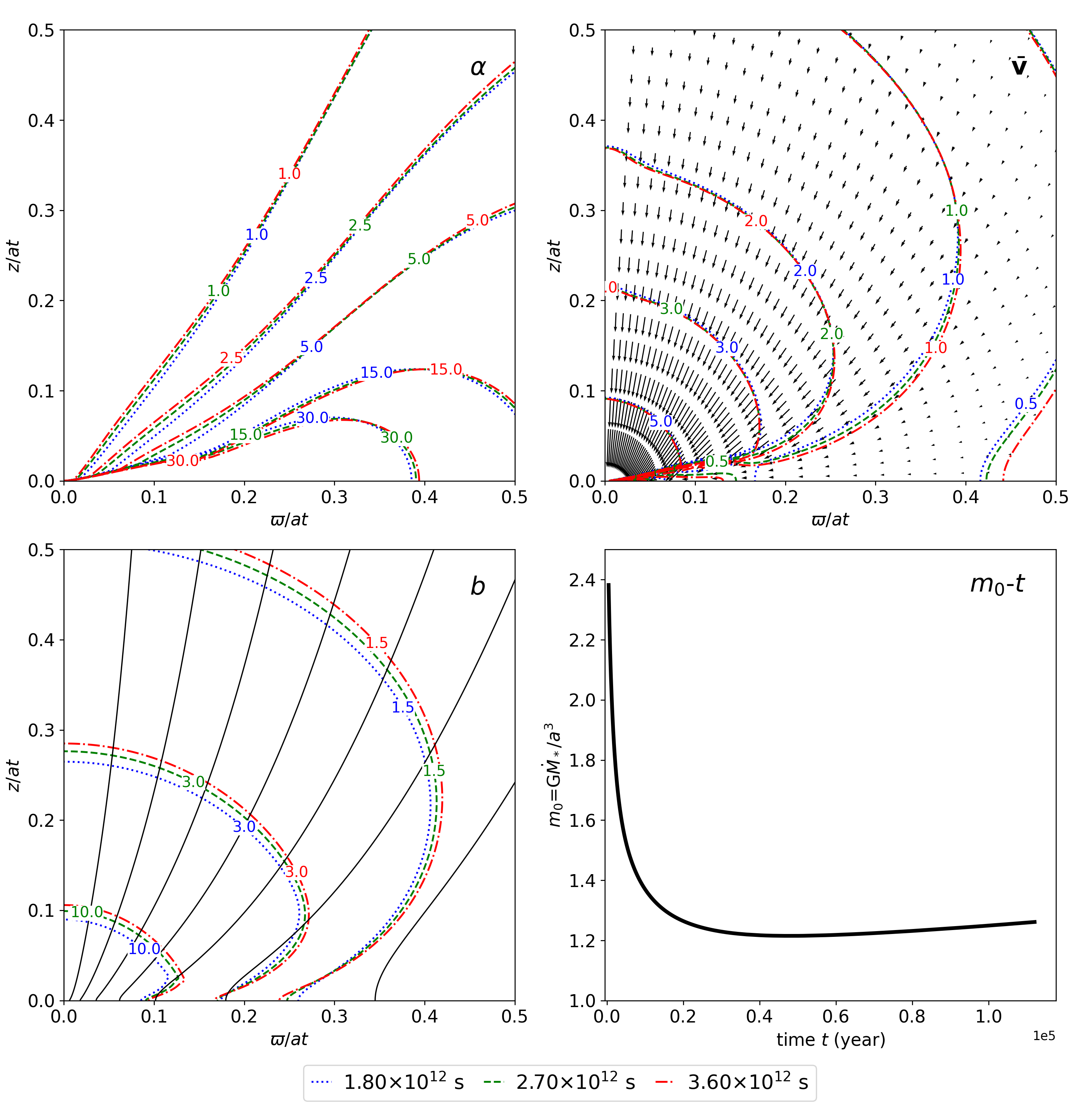}
    \caption{The results of four cases with parameters $n=1$ (top), $n=2$ (bottom), and $\eta=10^{18}$ (left) and $10^{19}\diffu$ (right).
    Each case is presented in four panels. The upper panels and the lower left panel respecively show the density, infall speed, and magnetic field strength at $t= (1.80\text{, }2.70\text{, }3.60)\times 10^{12}\second$ in dimensionless space (respectively as lines of blue dots, green dashes, and red dot-dashes). The arrows in the upper right panel show the velocity vectors at $t= 3.60\times 10^{12}\second$, and the solid lines in the lower left panel are magnetic field lines at the same time. The lower right panel shows the dimensionless mass infall rate $m_0 =\Dot{M}_* G/a^3 $ as a function of $t$.}
    \label{fig:n1n2_eta1819_selfsim}
\end{figure*}

We describe four reference simulations with $n=1$, $2$ and $\eta=10^{18}$, $10^{19}\diffu$ as an example to show the qualitative behavior of all cases.
Figure \ref{fig:n1n2_eta1819_selfsim} presents the reference cases in a manner similar to Figure 3 in \citet{allen2003}, using dimensionless coordinates.
Contour lines of the dimensionless field quantities $\alpha$, $v/a$, and $b$, at three representative times $t=(1.80$, $2.70$, $3.60)\times10^{12}\second$ are shown in the space of dimensionless cylindrical coordinates $\varpi/at$ and $z/at$.
The figure also shows vectors of velocity, magnetic field lines, and the evolution of $m_0(t)$.
The time-evolution of the contour lines of $\alpha$, $v/a$, and $b$ presents convergence in time, showing the approach to self-similarity. Self-similarity is not instantaneously achieved. The causes include the small differences between our set up and the ideal toroids (in our settings, the nonzero values of $\rin$ and $D^S$), the time it takes for the expansion wave of the inside-out collapse to reach scales large enough to become well resolved and interact with the initial toroid background, and the time needed for numerical transients to relax. Similar setup and numerical causes were already present in \citet{allen2003}. In this work, the explicit presence of physical resistivity introduces an additional time scale, the time necessary for diffusive effects to relax (see \S\ref{sec:diffusion_time} below).

The basics of the pseudodisk formation process operate similarly to the fiducial case of \citet{allen2003}. The magnetic field lines are pinched near the equatorial plane (lower left panel), dragged inwards by the gas infalling motion.
Field lines bend at the equator, with each field line having its own separate equatorial crossing at $z=0$.
The bent magnetic field lines produce an outwards-pointing magnetic tension force which decelerates the velocity of the infalling magnetized gas, shown (upper right panel) as shrinking dimensionless velocity contours near the equator, and deflected and shortened $\vv/a$ vectors.
This deceleration allows gas density to increase. The dimensionless density contours (upper left panel) present equatorial accumulation. This accumulation of gas partially supported (magnetically) against gravity forms the flat equatorial pseudodisk.
The magnetic field line configuration is not the same as in idealized MHD solutions. Because of explicit magnetic diffusivity, the dragging of field lines is at a rate slower than that of the gas infall, and the mass to flux ratio is not conserved. The field lines are strongly pinched, and they exert substantial magnetic tension force, but they do not significantly approach a radial configuration at the center.
We assess the effect of $\eta$ in these cases by comparing directly the case with $n=1$, $\eta=10^{18}\diffu$ (upper left panels) to the very similar case with $n=1$, $\eta=10^{18}\diffu$ (upper right panels). The case with lower diffusivity shows contours at different times which are comparatively much closer to each other, showing faster approach to self-similarity (see \S\ref{sec:diffusion_time} below). A few features are somewhat closer to the ideal MHD expectations for $\eta=10^{18}$ than for $\eta=10^{19}\diffu$.
The field lines are clearly more pinched, although they still remain very far from approaching a radial configuration. The velocity configurations are similar, but there are some differences, due to the more intense magnetic pinching. Away from the pseudodisk region, the vectors of velocity are less deflected and retain a configuration less deviated from radial. The thin equatorial region in which the velocity contours are strongly decelerated is even thinner for the less resistive case. In summary, the case $\eta=10^{18}\diffu$ shows a situation still very distinct from ideal MHD, but comparatively closer than for the qualitatively very similar $\eta=10^{19}\diffu$ case.
The case with $n=2$, $\eta=10^{19}\diffu$ (lower right panels) show in dimensionless units a case with larger $n=2$ than the $n=1$ value of the upper panels.
Comparison of the density contours $\alpha=30$ and $\alpha=15$ between these two figures shows increase of pseudodisk size and thickness with $n$. Velocity vector deflection is also stronger.
The approach to self-similarity by $t=3.6\times10^{12}\second$ is seen as less complete in the contours of magnetic intensity $b$ than for the corresponding $n=1$ case, and also in the time to approach the asymptotic value of $m_0$.

In agreement with the theoretical results \citep{shu1977,li1996b} and the simulations of \citet{allen2003}, the time-evolution of the central mass infall rate $m_0 =\Dot{M}_* G/a^3$ (lower right panel) asymptotically approaches a constant of order unity
in agreement with the expectations, and similar to the asymptotic $m_0(t)$ of \citet{allen2003}.
The numerical values are as follows.
For $n=1$, $\eta=10^{18}\diffu$, the value of $m_0(t)$ is seen to asymptote to $\sim1.14=0.91(1+H_0)$, with $M_*/(a^3t/G)=1.15=0.92(1+H_0)$.
For $n=1$, $\eta=10^{19}\diffu$, the asymptotic values are $m_0\sim1.15=0.92(1+H_0)$, $M_*/(a^3t/G)=1.16=0.93(1+H_0)$.
For $n=2$, $\eta=10^{18}\diffu$, the value of $m_0(t)$ asymptotes to $m_0=1.24=0.82(1+H_0)$, nearly the same as $M_*/(a^3t/G)=1.25=0.83(1+H_0)$.
For $n=2$, $\eta=10^{19}\diffu$, the value of $m_0(t)$ asymptotes to $m_0=1.27=0.85(1+H_0)$, nearly the same as $M_*/(a^3t/G)=1.28=0.85(1+H_0)$. These estimates of $m_0$ show an increase with $n=4H_0$, but with a coefficient somewhat smaller than the theoretical value of $0.975$ expected for small $H_0$ \citep{li1996b}. A similar reduced coefficient had been reported in \S3.2 of \citet{allen2003}.
The time necessary to approach and achieve the asymptotic value increases with $n$, with $m_0(t)$ still slowly growing by $t=1.2\times10^5\yr$.
These asymptotic values are similar to the theoretical expectation of $0.975 (1+H_0)$ of each case.

\begin{figure*}[htp]
    \includegraphics[width=\linewidth/2]{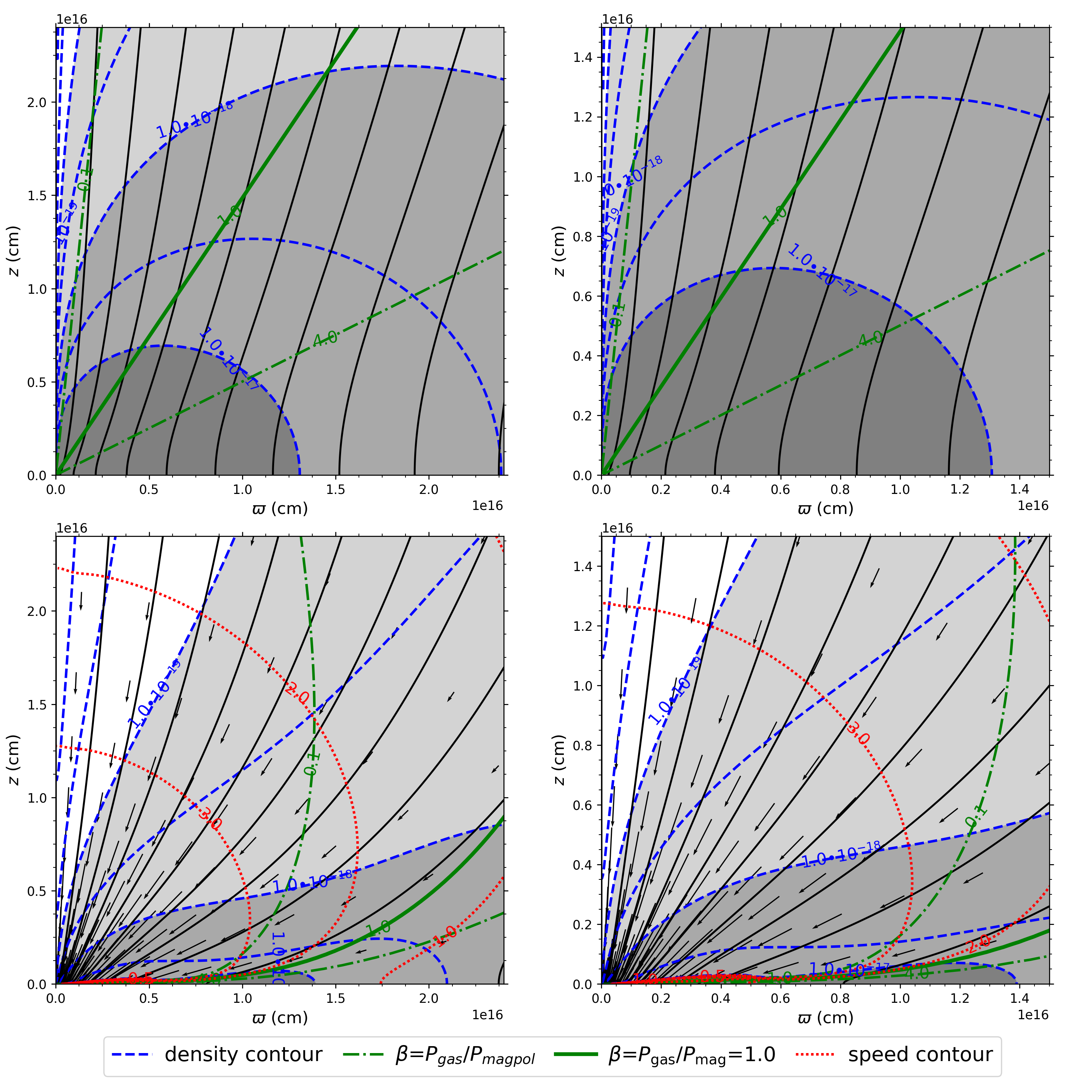}
    \includegraphics[width=\linewidth/2]{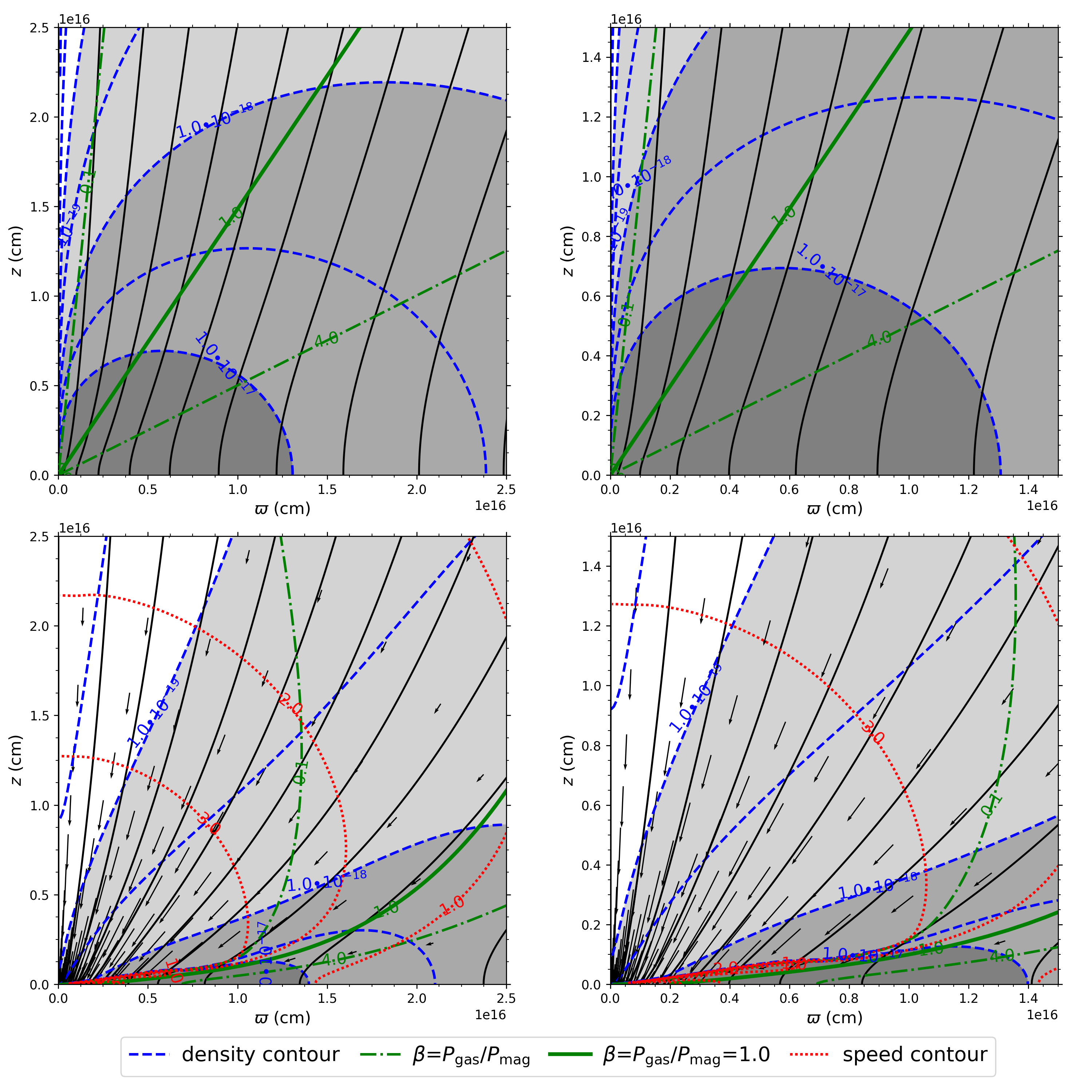}
    \includegraphics[width=\linewidth/2]{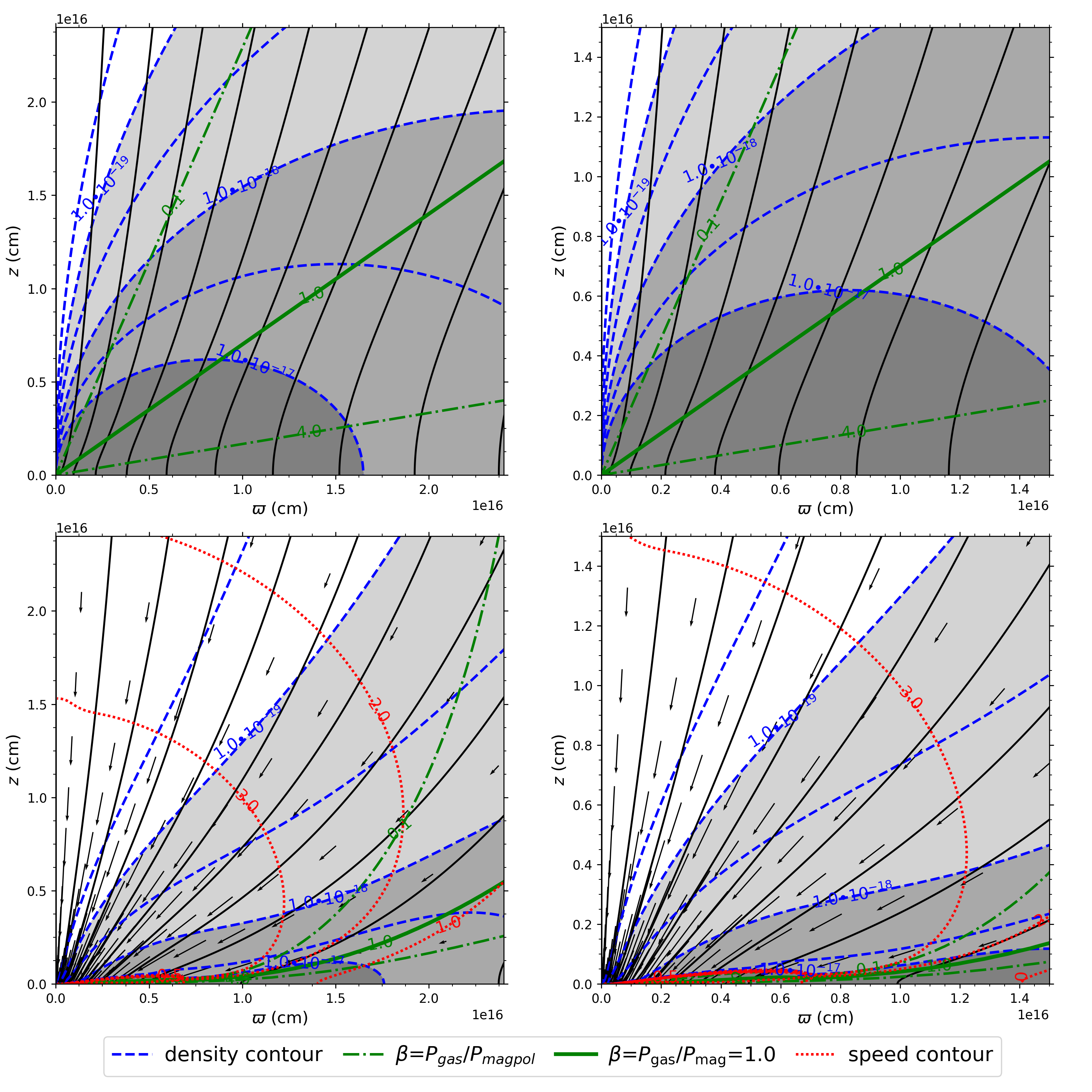}
    \includegraphics[width=\linewidth/2]{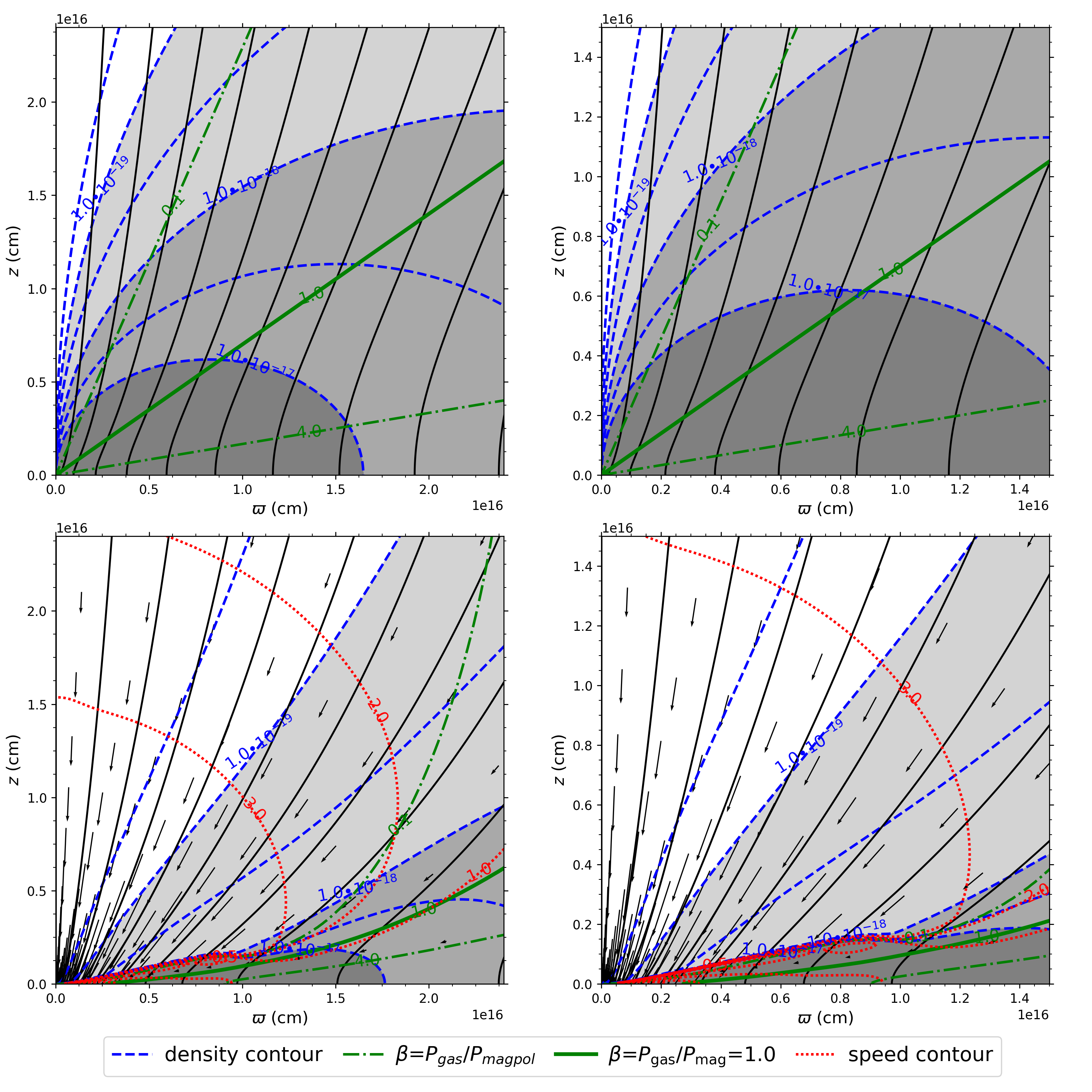}
    \caption{
    For the cases of Figure \ref{fig:n1n2_eta1819_selfsim}, a
    comparison of the initial setup at $t=0$ (upper panels) to the results at $t=3.6\times10^{12}\second=1.141 \times 10^{5}\yr$ (lower panels), for the case $n=1$, $\eta=10^{19}\diffu$. Results shown at two different scales: $2.5\times 10^{16}\cm$ (left) and $1.5\times 10^{16}\cm$ (right). Density contours in blue dashed lines highlighted by gray scale blocks, spaced two contours per decade up to $\rho=10^{-17}\massden$. Magnetic field lines in black solid. Red dotted lines show contours of $v/a$. Black arrows indicate velocity direction and relative magnitude. Plasma beta $\beta$ contours appear as green dash-dotted lines, with a solid green line marking the contour $\beta=1.0$.
    \label{fig:n1n2_eta1819_realspace}}
\end{figure*}

\begin{figure}
    \includegraphics[width=\linewidth]{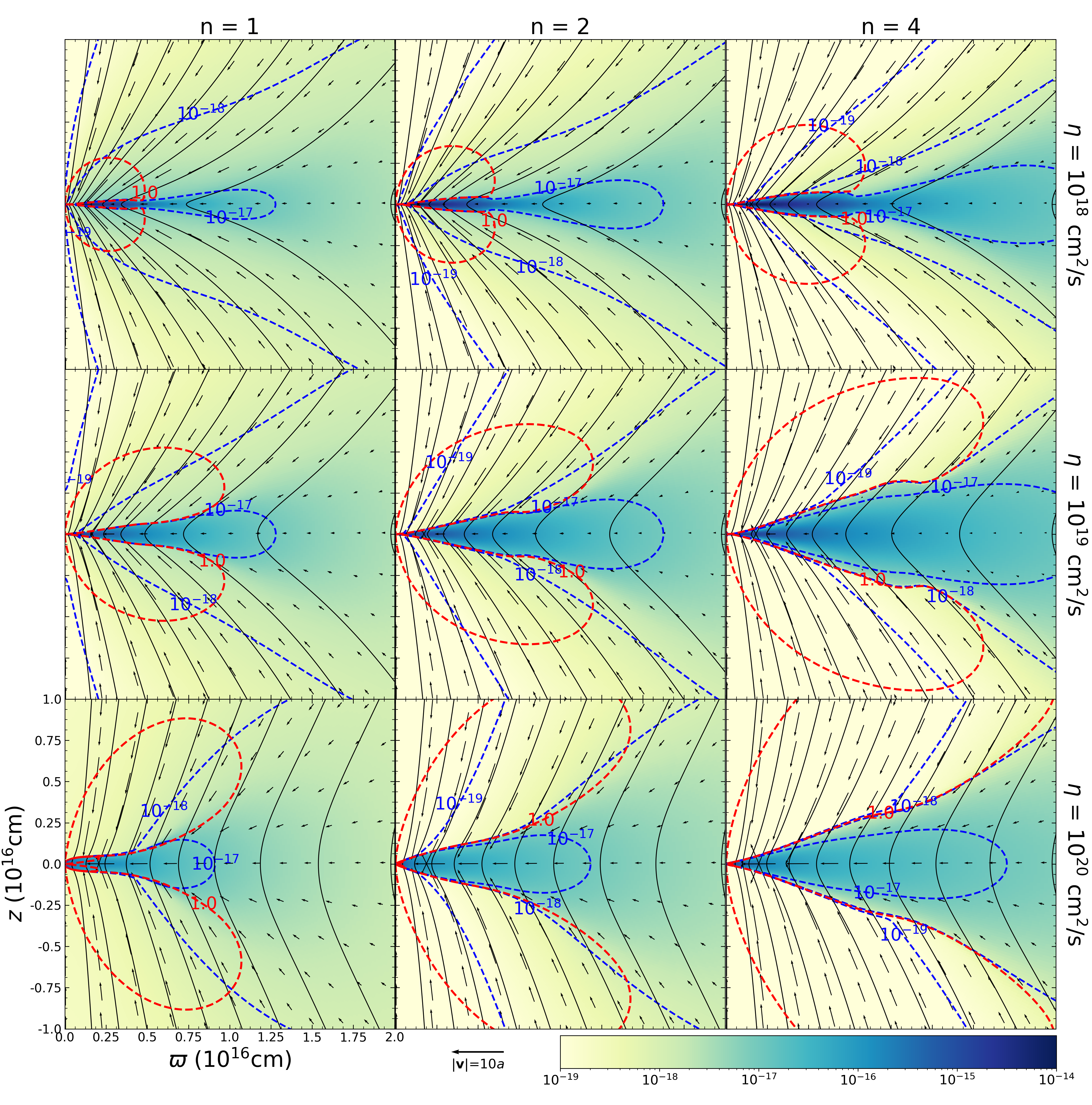}
    \includegraphics[width=\linewidth]{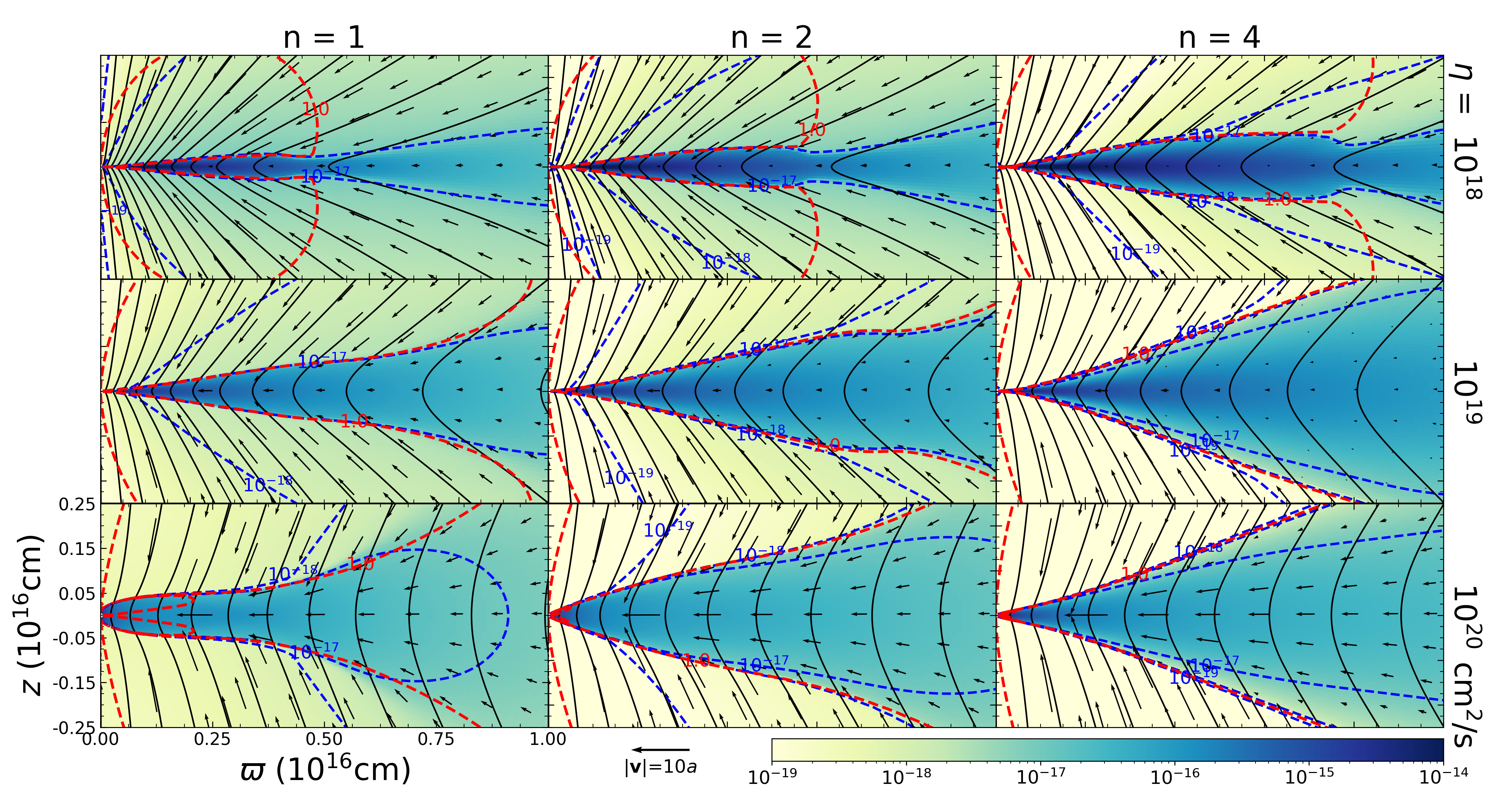}
    \includegraphics[width=\linewidth]{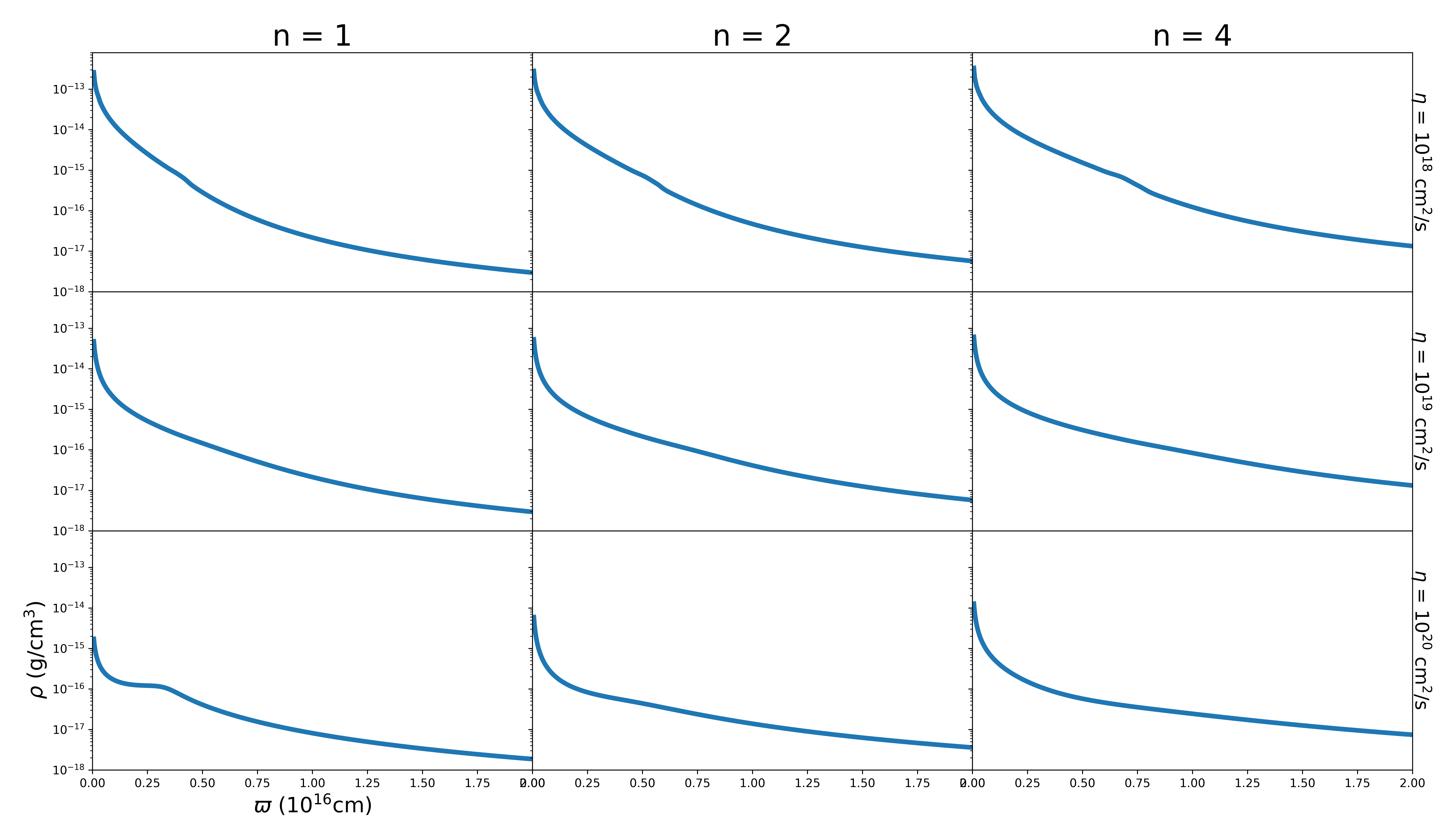}    
    \caption{Upper panel: colormap and three contours of density at $t=9.031\times10^4\yr$, together with field lines and velocity vectors. Columns: $n=1$, $2$, $4$. Rows: $\eta=10^{18}$, $10^{19}$, $10^{20}\diffu$.
    A contour line of $|v_\theta/a|=1$ is drawn in red dashes.
    The middle panel shows a close up near the midplane.
    The lower panel shows density as a function of $r$ at the midplane.
    }
    \label{fig:rho}
\end{figure}

Figure \ref{fig:n1n2_eta1819_realspace} presents in dimensional coordinate units the same four cases as in Figure \ref{fig:n1n2_eta1819_selfsim}. Each case is presented in four panels, analogous to Figure 4 in \citet{allen2003}. The top panels of each case show the initial conditions, and the bottom panels show the evolved results at time $t=3.6\times10^{12}\second=1.141\times10^5$ years.  The central masses are by this time equal to $M_*=4.91\times10^{32}\gram$, $M_*=5.02\times10^{32}\gram$, $M_*=5.38\times10^{32}\gram$, and $M_*=5.51\times10^{32}\gram$, respectively for \{$n=1$,$\eta=10^{18}\diffu$\}, \{$n=1$,$\eta=10^{19}\diffu$\}, \{$n=2$,$\eta=10^{18}\diffu$\}, and \{$n=2$,$\eta=10^{19}\diffu$\}. Contour lines, field lines, and velocity vectors are shown in dimensional cylindrical coordinates $\varpi$ and $z$ (in $\cmNSP$). Pseudodisk formation takes place analogously to the results reported in \citet{allen2003}: the regions where plasma beta $\beta = P_\text{gas}/P_\text{mag}<1.0$ expand over time from their initial polar zones to occupy a much larger volume, which however still excludes the equatorial zones, occupied by the growing pseudodisk.
In the scale of $2.5\times10^{16}\cm$, the density contours illustrating the pseudodisks are slightly thicker than for the corresponding case in \citet{allen2003}, especially in their outer parts. This difference is due to the presence of explicit resistivity.
The contour $\rho=10^{-17}\massden$ is now well resolved in $z$ in all cases.

Figure \ref{fig:rho} provides a comparison of pseudodisk shapes for our nine models at a fixed time, in dimensional units.
The two top parts of the figure show density contours and field lines for three values of $\eta$, increasing from top to bottom, and three values of $n$ increasing from left to right, at the same given time $t=9.031\times10^4\yr=2.85\times10^{12}\second$.
The density contour $\rho>10^{-17}\massden$ can illustrate the pseudodisk shape, Decreasing the diffusivity from $\eta=10^{20}\diffu$ to $\eta=10^{18}\diffu$ makes these contours thinner in $z$, while larger in $\varpi$, showing faster pseudodisk growth. Increasing the value of $n$ has produced larger density contours in both dimensions, indicating its effect on pseudodisk size.
Field lines are not radial, and strongly deviate from the $\eta=0$ ideal MHD model for pseudodisks. The field lines at the equator are vertical. Field line shapes become less bent and more vertical as $\eta$ increases and infall becomes more decoupled from the magnetic field. In all cases matter can fall to the top and bottom surfaces of the pseudodisk directly. All nine panels of Figure \ref{fig:rho} indicate shocks at the pseudodisk surface, as abrupt, discontinuous changes of density levels as shown in contours and in the colormaps.
The lower part of Figure \ref{fig:rho} shows equatorial values of density as a function of radius. They are typically in the range $<10^{-14}\massden$, and rise up to a maximum of no more than a few times $10^{-13}\massden$, always well below a value range $\gtrsim 3 \times 10^{-12}\massden$ (corresponding to $10^{12}\numden$) for which rapid increases are expected \citep{zhao2018}.

Analysis of the velocity vectors in Figure \ref{fig:rho} reveals that the magnetized infall in our models is not simply free fall. The vectors illustrate four distinct phases of this process: deflection, shock, redirection, and magnetic slow-down.

Well above the pseudodisk midplane, the velocity infall vectors deviate from radial paths. They are \textit{deflected towards the midplane} by magnetic forces and self-gravity, feeding the pseudodisk from its upper and lower surfaces.

At these surfaces, an \textit{oblique shock} forms, evident in Figure \ref{fig:rho} as a discontinuous change in density over large portions of the pseudodisk surface. The velocity vector at this oblique shock has both normal and tangential components. The normal component undergoes a discontinuity and is strongly reduced in the post-shock region, while the tangential component changes continuously. Consequently, the velocity direction is \textit{redirected} towards the center.

This redirection is illustrated in Figure \ref{fig:rho} by the post-shock vectors and the behavior of the dashed contour line of $|v_\theta/a|$ at the oblique shock. The $v_\theta$ component, approximately normal to portions of the shock surface, is discontinuously reduced as it transitions from supersonic to subsonic, causing the contour line to overlap with the shock surfaces at these points.

After traversing the oblique shocks, matter enters the denser regions of the pseudodisk, where magnetic tension increasingly counteracts gravity. The resulting \textit{magnetic slow-down} reduces the velocity vectors below the free-fall value. Section \ref{sec:ratios} is dedicated to examining the details of this crucial part of the accretion process.

\section{Discussion} \label{sec:discussion}
In this section we show and discuss the force and velocity ratios produced in resistive toroid collapse (\S\ref{sec:ratios}), and the role of $\eta$ in the timescale of approach to self-similarity (\S\ref{sec:diffusion_time}). We then discuss the implications of this work in \S\ref{sec:implications}.

\subsection{Force and velocity ratios}\label{sec:ratios}
In the study of collapse, the relative importance of magnetism and gravity can take different forms.
In the case of our collapsing singular isothermal toroids, they have been reported in \S\ref{sec:results} to feature regions of partial magnetic support, not strong enough to prevent or reverse collapse, but strong enough to present local enhancements of density.
In this \S\ref{sec:ratios} we quantify that report more precisely by showing the ratios of the radial components of the magnetic vs gravity force. We also compare the effects of magnetic fields to the gravity and hydrodynamic physics by taking the ratios of the radial infall velocity component $v_r$ in these collapsing toroids to three relevant reference velocities characteristic of hydrodynamic flows: the sound speed $a$, the free fall speed $v_\text{ff}=(2 G M_*/r)^{1/2}$, and the solution $v_\text{SIS}(r)$ of the singular isothermal sphere collapse \citep{shu1977}.

\begin{figure}
    \includegraphics[width=\linewidth]{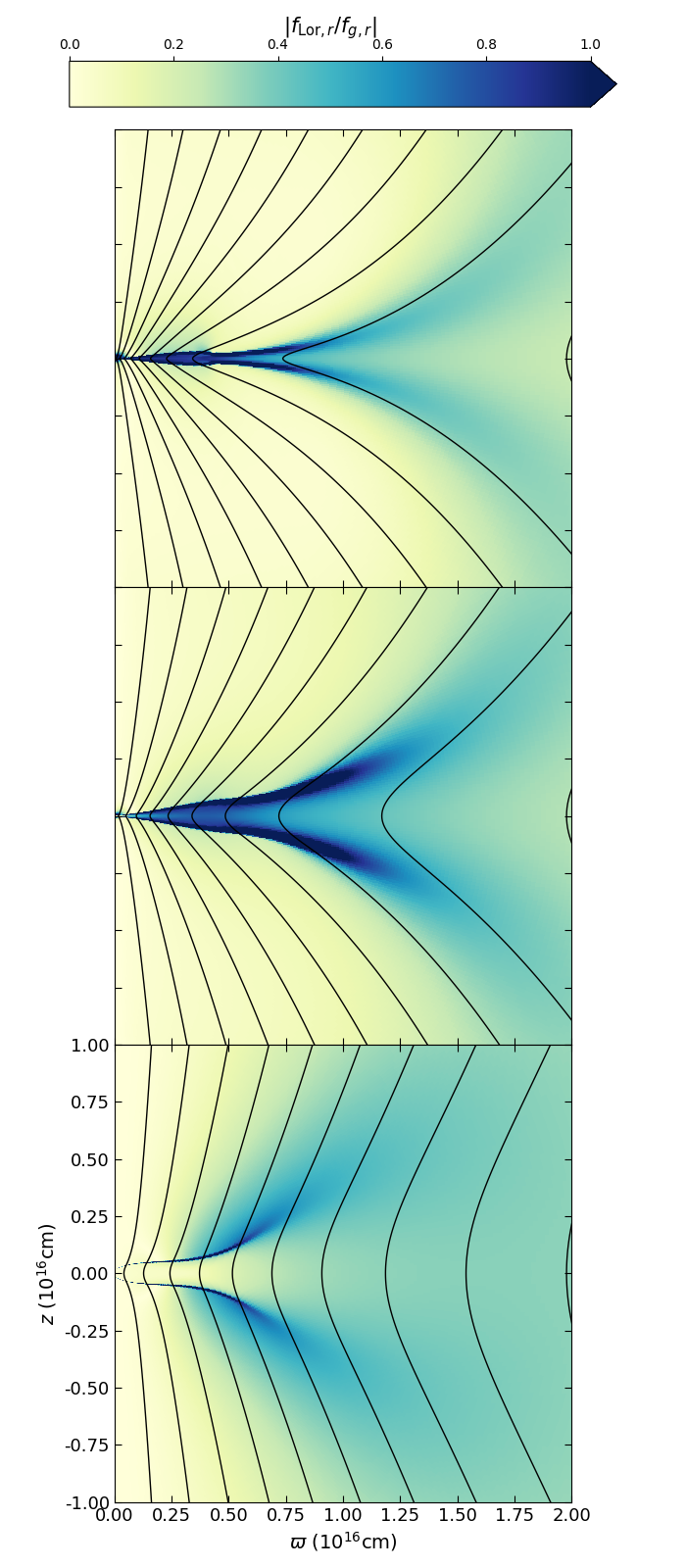}
    \caption{Colormap of the ratio of the $r$ component of Lorentz force to that of the gravitational force $|f_{\mathrm{Lor},r}/f_{g,r}|$ at $t=2.85\times 10^{12}\second$ for $n=1$ and three values of $\eta = 10^{18}$, $10^{19}$, and $10^{20}\diffu$ from top to bottom.
    }
    \label{fig:n1_FlorrFgr_2e16}
\end{figure}

In Figure \ref{fig:n1_FlorrFgr_2e16} we compare the relative roles of magnetism and gravity by showing colormaps of the ratio $|\lfrac{f_{\mathrm{Lor},r}}{f_{g,r}}|$ of radial force components
for $n=1$ and the three different values of Ohmic diffusivity, at time $t=2.85\times 10^{12}\second$. At this time, the central mass is $M_*=3.88$, $3.99$, and $4.68\times10^{32}\gram$, respectively for $\eta = 10^{18}$, $10^{19}$, and $10^{20}\diffu$, large enough to become the dominant term of the gravity force in the inner parts.
We obtain that in all three cases, the local maximum of the force ratio $|f_{\mathrm{Lor},r}/f_{g,r}|$ delineates the outlines of the pseudodisk. At large heights $|z|$ above and below the pseudodisk, the current density is small, reducing the magnetic force to a small value.
Very close to the pseudodisk equator, the density $\rho$ is large, relatively decreasing the radial Lorentz force per unit mass
below its value at the pseudodisk surface.
This equatorial decrease is more significant for the case $10^{20}\diffu$, whose less well-coupled magnetic field lines have smaller equatorial bending within the pseudodisk. For the other two values of $\eta$ the force ratio stays large even at the equator.
\citet{shu1997_isopedic} explored the ratio of radial magnetic forces to gravity, obtaining a fundamental concept of magnetic dilution of gravity, which in the ideal MHD limit leads to a simple function of a constant mass-to-flux value $\lambda$. This is generalized in our nonideal MHD study to a function of $\eta$ and position within the pseudodisk.

\begin{figure*}
    \includegraphics[width=\linewidth]{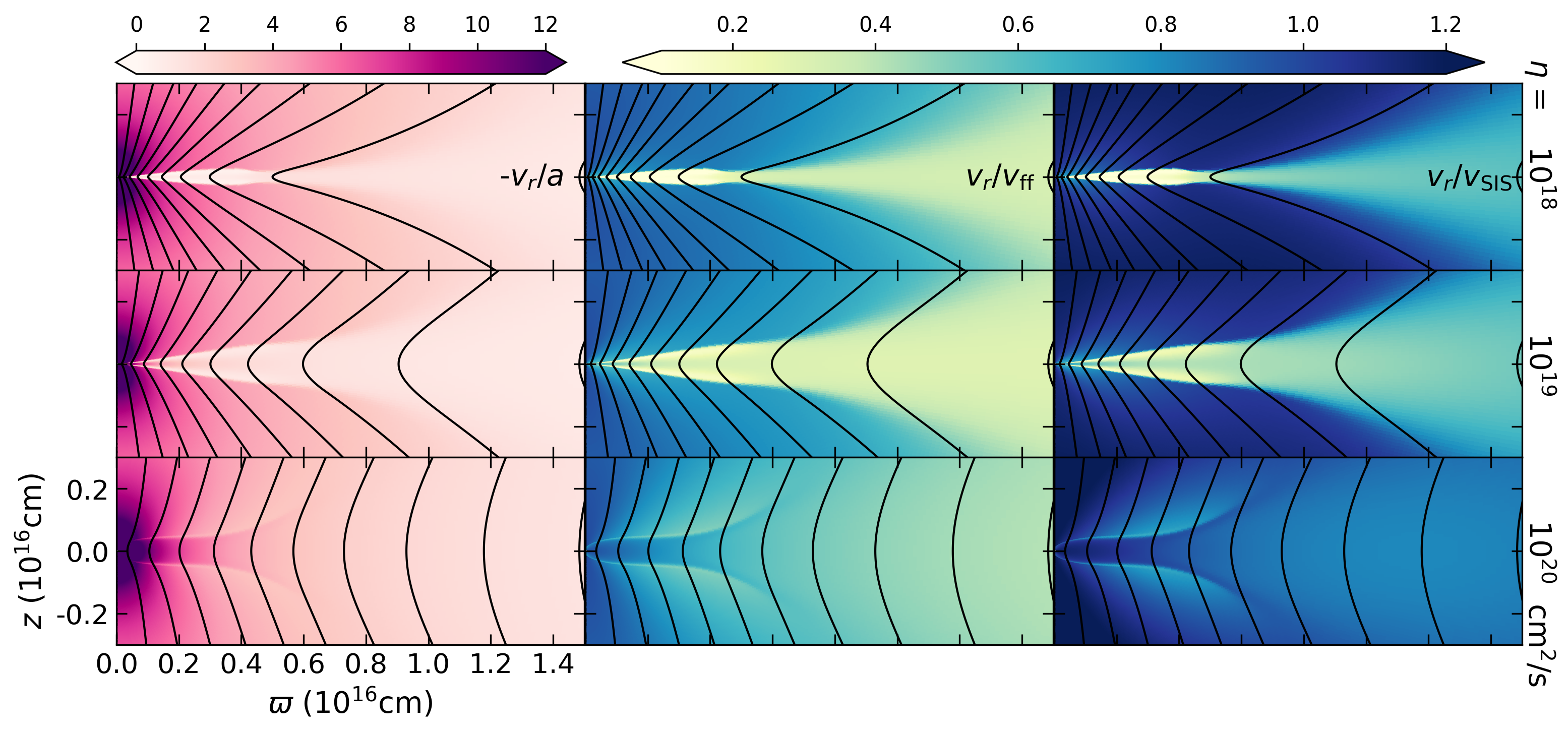}
    \caption{Colormaps of velocity ratios. From left to to right, the dimensionless radial velocity $-\bar{v}=-v_r/a$ (top), the ratio of radial velocity to free fall $v_r/v_\text{ff}$ (middle), and the velocity ratio $v_r/v_\text{SIS}$. Data taken at $t=2.85\times 10^{12}\second$ for $n=1$ and three values of $\eta = 10^{18}$, $10^{19}$, and $10^{20}\diffu$ from top to bottom. The value of $v_\text{SIS}(r)$ is obtained by solving the ODE system of \citet{shu1977} with a parameter $A=2.00001$.}
    \label{fig:vr_20e15}
\end{figure*}

Large values of the force ratio $|f_{\mathrm{Lor}, r}/f_{g,r}|$ provide magnetic support against gravity, and have consequences on the velocity ratios, shown in Figure \ref{fig:vr_20e15} for the same cases and times of Figure \ref{fig:n1_FlorrFgr_2e16}.
The left column shows the radial component of the dimensionless velocity, $-v_r/a$.
The speed of the gas-infalling motion decreases in regions where the Lorentz force is comparable with the gravitational force. The colormaps of force and velocity ratios become correlated, indicating the role of magnetic support in pseudodisk formation.
However, for $\eta=10^{20}\diffu$, weak coupling allows clearly supersonic infall in the inner equatorial region.

The midle column compares $v_r$ to the free-fall velocity $v_\text{ff}$.
Magnetic support is able to keep the infall velocity within the pseudodisk well below $v_\text{ff}$. The local minima of $v_r/v_\text{ff}$ are located at the pseudodisk top and bottom surfaces, overlapping with the local maxima of $|f_{\mathrm{Lor},r}/f_{g,r}|$. 
By contrast, in regions far above and below the pseudodisk, where the magnetic support ratio is small, $v_r$ can become close to the free-fall velocity $v_\text{ff}$, and the ratio $v_r/v_\text{ff}$ achieves its local maxima $\approx 1$.
For $\eta=10^{20}\diffu$, weak coupling allows infall velocities not so far below free fall also in certain parts of the equator.

The right column of Figure \ref{fig:vr_20e15} shows the ratio of $v_r$ to the radial infall velocity $v_\text{SIS}$ in the self-similar collapse of a hydrodynamic, nonmagnetized singular isothermal sphere \citep{shu1977}.
This panel aims to compare the resistive, magnetized collapse to the SIS spherically symmetric model of hydrodynamic collapse.
The pseudodisk shows regions of relatively small $v_r/v_\text{SIS}$, with local minima located close to the local maxima of $|f_{\mathrm{Lor},r}/f_{g,r}|$, patterns already noticed in the $v_r/v_\text{ff}$ panel. Typical values of the ratio within the pseudodisk range from $0.1$ to $0.5$, and show stronger $\theta$ dependence in the inner pseudodisk scale ($\varpi\lesssim0.5\times10^{16}\cm$).
(However, for $\eta=10^{20}\diffu$, weak coupling allows infall velocities not far below $v_\text{SIS}$ along the equator.)
Above and below the pseudodisk, where magnetic support against gravity is minimal, infall is near nonmagnetized spherical infall conditions, and the ratio $v_r/v_\text{SIS}$ can approach or exceed unity, such in axial regions where $v_r\approx v_\text{ff}\gtrsim v_\text{SIS}$.


\begin{figure}
    \includegraphics[width=\linewidth]{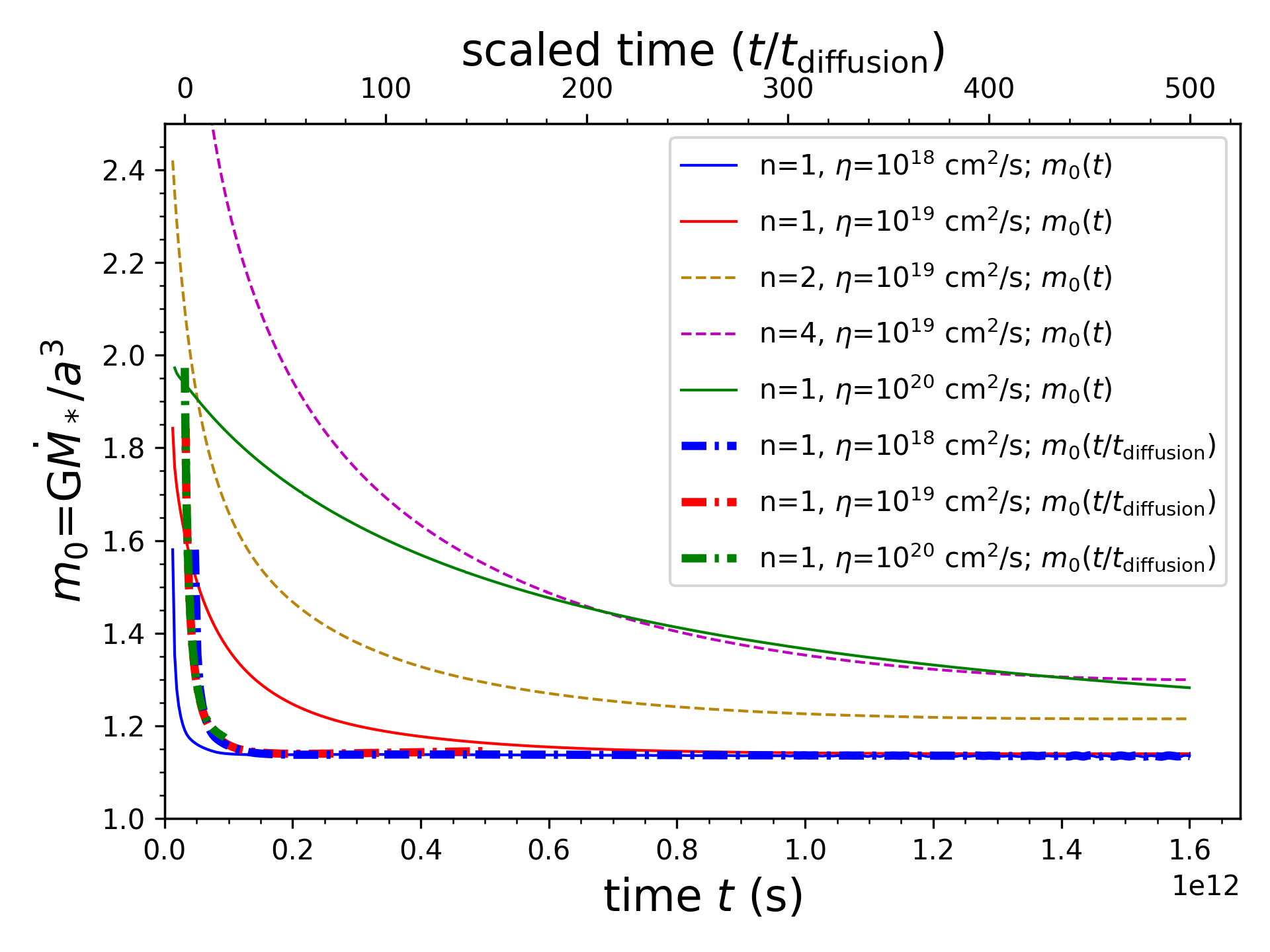}
    \includegraphics[width=1.08\linewidth]{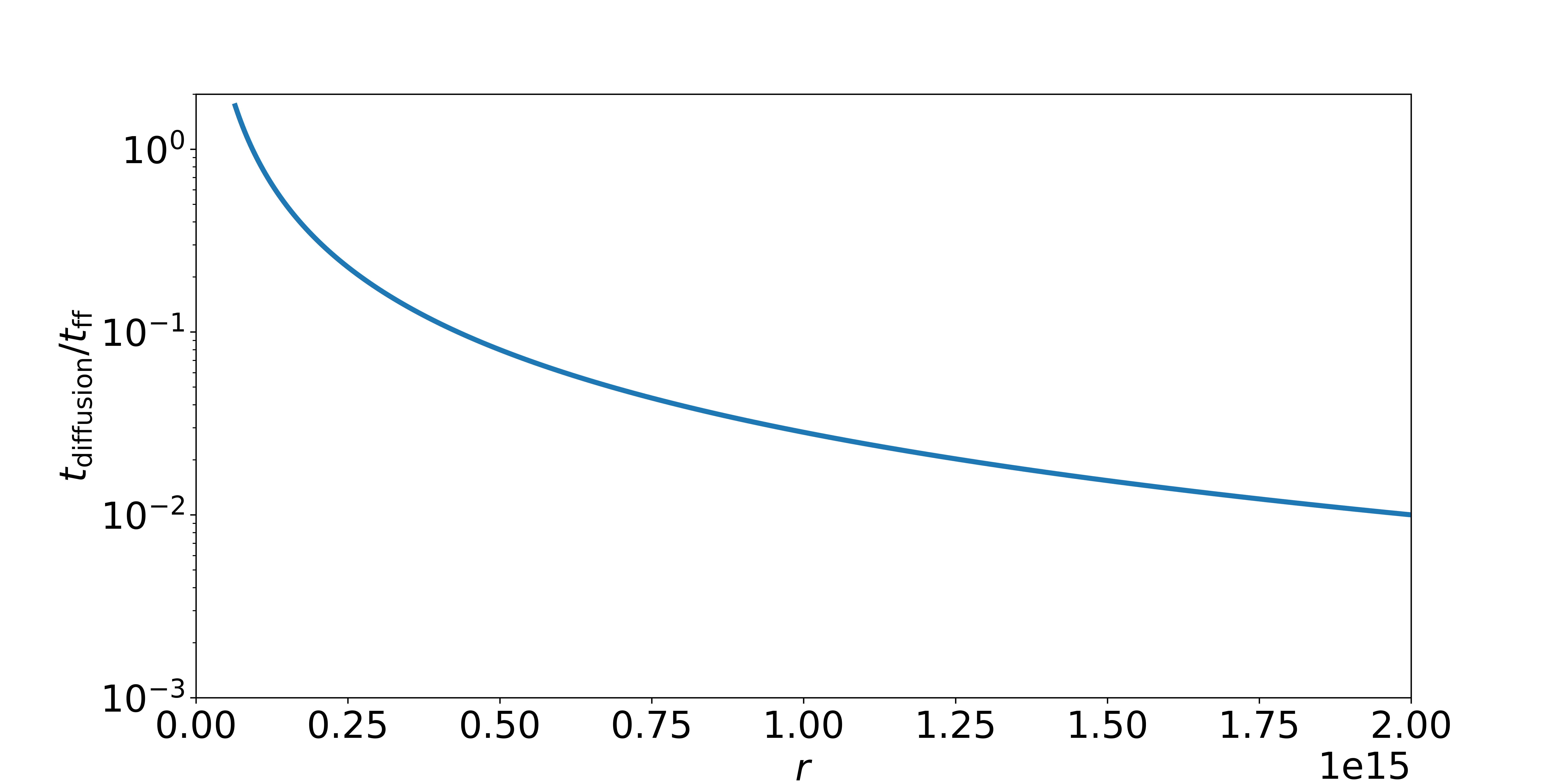}
    \caption{Top: evolution of the dimensionless mass growth rate $m_0$ in time. Dashed and solid lines show $m_0$ as a function of dimensional time $t$ (in seconds, lower horizontal axis). Thicker, dash-dotted lines show $m_0$ as a function of the dimensionless time $t/t_\mathrm{diffusion}$ (shown on the upper horizontal axis of scaled time).
    Bottom: Ratio of $t_\text{diffusion}/t_\text{ff}$ at an early time in the simulation, for the case $n=1$, $\eta=10^{18}\diffu$.
    }
    \label{fig:m0t_sct}
\end{figure}

\subsection{Ohmic diffusivity and self-similarity}\label{sec:diffusion_time}
Comparing dimensionless contour lines for different values of $\eta$
(upper panels of Figure \ref{fig:n1n2_eta1819_selfsim}) shows that larger $\eta$ values have a slower approach to self-similarity.
Dimensional analysis shows that the existence of $\eta$ as a parameter introduces natural scales of length and time
\begin{eqnarray}
    r_\mathrm{diffusion} &\equiv& \eta /a \label{rdiffusion}\\
    t_\mathrm{diffusion} &\equiv& {\eta}/{a^{2}}, \label{tdiffusion}
\end{eqnarray}
breaking the strict self-similar conditions.
Self-similarity can be recovered when the natural time scale of diffusion $t_\mathrm{diffusion}$ is much smaller than the time since point mass formation $t$, which in our settings is similar to the the accretion time scale.
This is because apart from a brief initial transient, the accretion rate $\dot{M}_{*}$ is always within an order of unity of $a^3/G$, and therefore the accretion time scale $t_\mathrm{acc} =M_{*}/\dot{M}_{*}$ is always $t_\mathrm{acc}\approx t$.
The Ohmic time scale of \citet{shu2006}, $t_\mathrm{Ohm} \equiv {\eta ^{3}}/{4G^{2}M_{*}^{2}}$ corresponds to a situation close to free-fall to a central mass $M_*$, conditions different from the present settings that are frequently out of free fall, and have a time-dependent $M_{*}$ which approximately fulfills the formula $\dot{M}_{*}=0.975(1+H_0)a^3/G$ of \citet{li1996b}.

The time evolution of the dimensionless mass infall rate $m_0(t)$ (top panel of Figure \ref{fig:m0t_sct}) demonstrates a clear dependence of the approach to self-similar state on $t_\mathrm{diffusion}$. At late times, each curve $m_0(t)$ approaches an asymptotic value, depending on $n$ (dashed and solid lines in Figure \ref{fig:m0t_sct}, showing $m_0(t)$). For larger $\eta$, the approach to the asymptote is always slowed down, with a time scale that can be quantified.
Rescaling the curves to $m_{0}(t/t_\mathrm{diffusion})$ (thicker, dash-dotted lines Figure \ref{fig:m0t_sct}, showing scaled time for $n=1$) results in lines that nearly overlap each other and reach asymptotic state at late times. (The case $10^{20}\diffu$ falls on the same scaled curve, although the asymptotic state is not yet completely achieved within simulation time.)
The time to reach within 10\% of the asymptotic state is measured to be $\sim10 t_\mathrm{diffusion}$ for all the three lines ($9 t_\mathrm{diffusion}$ for $\eta=10^{18}\diffu$, 7 for $\eta=10^{19}\diffu$, and $\gtrsim6$ for $\eta=10^{20}\diffu$).
This demonstrates a time scale $\propto t_\mathrm{diffusion}$.

The bottom panel of Figure \ref{fig:m0t_sct} shows an analytical estimate of the ratio of $t_\text{diffusion}$ to a free fall time $t_\text{ff}=r^{3/2} (2 G M_{*})^{-1/2}$, computed for the mass at an early time $t=3.2\times10^9\second>t_\text{diffusion}=2.5\times10^9\second$. It is remarkable that the typical value of the time scale ratio is that the diffusive time is faster than the free fall time.
The early processes of pseudodisk formation take place at the same time as the early times of central mass accumulation, when $M_*$ and $v_\text{ff}$ are still small.

The values of magnetic diffusivity explored in this work can be compared with values in recent literature.
Equation (10) in \citet{TsukamotoPP7} estimates the ambipolar diffusivity for various ranges of density. Applying it to our $\rho \sim 10^{-18}$--$10^{-13}\massden$ (Figure \ref{fig:rho}c) leads to a diffusivity in the order of $2\times 10^{18}$--$2\times 10^{19}\diffu$, covering very well most of our range of intensity of the magnetic diffusivity. However, approximate values for Ohmic diffusivity in their Equation (9) lead to values in the range $10^{11}$--$10^{16}\diffu$, far below those reported in this study. \citet{zhao2018} had already noticed that for number densities in the range $\lesssim 10^{10}\numden$, ambipolar resistivity can be orders of magnitude larger than Ohmic resistivity.
Our simulations of disk formation utilize magnetic diffusivities in the order of $\geq10^{17}$--$10^{18}\diffu$ and higher, comfortably larger than the numerical diffusivity corresponding to the fine-resolution of our spherical grids. The values $>10^{18}\diffu$ are also sufficiently large to prevent the numerical oscillations reported in Section \ref{sec:toroid_collapse}, and in \citet{allen2003} \S3.2.
Our models utilize a uniform magnetic diffusivity. This can be compared with models having distributions of resistivity within the pseudodisk regions.
Within our ranges of densities within the pseudodisk, the AD magnetic diffusivity of Equation (10) in \citet{TsukamotoPP7} is not far from a uniform value $2\times 10^{18}\diffu$, and it would coincide with our estimates of magnetic diffusivity.
\citet{vaisala2023} utilizes a resistivity proportional to $\rho^{1/2}$, which may be used as a comparison in future work.

\subsection{Implications}\label{sec:implications}
This nonideal MHD revisit of \citet{allen2003} has confirmed the major features of collapse but demonstrated the need to revise the pseudodisk structure. The inside-out expansion wave of the collapse is essential as in hydrodynamic theory, and the predictions of ideal MHD for the growth rate of the central mass have also been confirmed.

However, unlike in ideal MHD, magnetic field lines do not approach a radial configuration when magnetic diffusivity is present. This critical qualitative result is expected to hold true for Ohmic diffusivity and any diffusive magnetic microphysical model. The change in field line shapes alters how matter accretes onto the pseudodisk, regularly producing oblique shocks at its surfaces. The strength and location of these shocks would depend on both the magnitude and spatial distribution of magnetic diffusivity.

We remark on a theoretical implication on how these shocks are formed. In both hemispheres, the flow falls in from large heights, with streamlines that are deflected towards the midplane. The oblique shocks we have noticed in Figure \ref{fig:rho} are not however the result of a direct convergence of two infalling streams from $\pm z$. Instead, each stream has its own oblique shock at two well-separated upper and lower shock surfaces. Each stream smoothly and separately transitions to the equatorial infall further inside.
However, convergence shocks at or near the midplane remain possible for a different configuration or parameters, e.g., lower diffusivity, which may allow flows within a much thinner pseudodisk to converge.

We now mention astrophysical features and processes that have implications for observation.

\paragraph{Magnetic force estimates}
In section \S\ref{sec:ratios} we showed a concrete example of how the ratio of infall slowdown $v_r/v_\text{ff}$ is correlated with the magnetic force ratio $|f_{\mathrm{Lor},r}/f_{g,r}|$. These ratios have applications to methods such as those developed and applied in \citet{koch2012}, \citet{aso2015}, \citet{sai2022}, and \citet{yen2023}. These methods combine observational information about magnetic field line structure, mass distribution, and infall kinematics to deduce magnetic field intensity estimates from a gravitational to magnetic forces comparison. Our concrete examples, through synthetic observation and direct comparison of simulated forces, can help firm up the theoretical bases of these methods.

\paragraph{Shock heating and its implications}
The oblique shocks at the pseudodisk surfaces can undergo shock heating at a place with large density gradients. It is possible that localized high temperatures would be sufficient to enhance reaction rates, and start the activation of astrochemical processes, perhaps including COM reactions and CO desorption.

The picture of these oblique shocks has the potential of becoming more complex in the presence of winds launched from pseudodisks (e.g.\ recent simulations in \citealt{basu2024}).

\vskip 3ex

Many observable core collapse objects can be modeled in the manner of this study. Here we review a few of them.

\paragraph{B335}
This Class 0 protostellar source has measurements of highly pinched magnetic field on the $1000\au$--$50\au$ scale \citep{maury2018}. Infalling motion in the $3000$--$10\au$ scales has been detected \citep[e.g.][]{zhou1993,saito1999,yen2015,bjerkeli2019}.
There are no clear signs of Keplerian rotation in the $\gtrsim10\au$ scale \citep{yen2015,bjerkeli2019}, setting an upper bound to the Keplerian disk size.
The structure of the envelope has been reported. The large scale 
\citep{zhou1993,saito1999,kurono2013} is identifiable with an SIS density profile \citep{shu1977}. The flattened structure is seen at a smaller scale \citep[e.g.,][]{yen2010}, with evidence of the position of the expansion wave at $r\sim4000\au$ \citep{kurono2013,evans2015,evans2023}.
The pinched hourglass-shaped magnetic field, together with the envelope structure and infalling motions, indicate that this is a place for application of the models and theory of this work.

\paragraph{HH 211} 
This very young Class 0 system shows a pinched poloidal field magnetic structure \citep{lee2009,lee2014_hh211,lee2019}.
A flattened structure is observed in the large scale \citep{lee2014_hh211}.
Because of the extreme youth of HH 211, the protostar is deeply embedded.
The envelope is heavily infalling and resembles \citep{lee2019} a stage of the SIS solution \citep{shu1977}.
This is close to our situation for $n\lesssim1$ for very early times.

\paragraph{HH 212}
This Class 0 system also has specific features for which these simulations can be applied.
Its envelope has an infall that is not radial, but deflected towards the midplane \citep{lee2014_hh212}, similar to the model of this work as illustrated in the velocity vectors of Figure \ref{fig:rho}, and to the field line structure in quasistatic toroids.
Complex organic molecules (COMs) are detected in a scale from $\sim40\au$ to $\sim10\au$ \citep{lee2017}.
Shocks may facilitate COMs to form by raising the temperature.

\paragraph{L1157}
This source presents an hourglass-shaped magnetic field \citep{stephens2013}, and infall kinematics \citep{chiang2010}. These properties have been described \citep{looney2007} as a Class-0 detection of the kind of flattened envelope structure that derives into pseudodisks.
Its morphology and kinematics mark it as a place for possible application of this work.

\section{Summary} \label{sec:summary}
The magnetized singular isothermal toroids \citep{li1996b} represent a late prestellar phase in molecular cloud cores. In this work, we revisit the process of their collapse (studied in \citealt{allen2003} using a code without explicit resistivity) by adding the explicit effects of Ohmic diffusion, leading to a more realistic depiction of the collapse and also obtaining numerical advantages for the computers of today. We consider the collapse of toroids with $n=4H_0=1$, $2$ and $4$, and apply three levels of uniform Ohmic diffusivity, $\eta=10^{18}$, $10^{19}$, and $10^{20}\diffu$.

The main basic qualitative features of the previous study are retained. The toroids collapse from the inside out \citep[as in the HD solutions of][]{shu1977}, with a wave of collapse moving outwards at essentially the isothermal sound speed $a$. 
At the center, a central point mass is generated with a growth rate $\Dot{M}\sim0.975(1+H_0) (a^3/G)$, consistent with the results of \citet{allen2003} and theoretical predictions in \citet{li1996b}. There are differences due to the presence of explicit diffusivity.
Self-similarity is approached after the timescales of the expansion wave $r/a$ and of the diffusive time scale $\eta/a^2$ have sufficiently elapsed.

As in \citet{allen2003}, the collapse forms pseudodisks, equatorial inner structures featuring partial magnetic support by the magnetic tension of bent field lines \citep{galli1993_i,galli1993_ii}.
Density contours (Figure \ref{fig:rho}) illustrate shape and size of the pseudodisks. Decreasing $\eta$ produces more pinched field lines threading pseudodisks more extended in $\varpi$. Increasing $H_0$ produces larger and thicker density contours.

The presence of magnetic diffusivity makes the field lines less radial than in the ideal MHD limit, helping matter to fall directly onto the pseudodisk surfaces. That infall produces shocks at the pseudodisk surface.
Shock heating might reach temperatures high enough to activate astrochemical effects in molecular chemistry.

Structure growth, force ratios, velocity infall, and field line configurations depend on resistivity.
For higher values of magnetic diffusivity, the field line configurations become less pinched, allowing the infall velocities to approach a larger fraction of the free-fall speed.
This diffusive control of the infall slowdown rate connects microphysics to an important observable quantity influencing mass accretion.
The astrophysical consequences will be used to make observable predictions about structures within collapsing envelopes and can be generalized to nonuniform magnetic diffusivity, including other nonideal MHD processes.

\begin{acknowledgments}

The authors acknowledge support for the CHARMS Project from the Institute of Astronomy and Astrophysics, Academia Sinica (ASIAA), the Academia Sinica grant AS-IAIA-114-M01, and the National Science and Technology Council (NSTC) in Taiwan through grants 112-2112-M-001-030, 113-2112-M-001-008, and 113-2927-I-001-513-. The authors thank the National Center for High-performance Computing (NCHC) of National Applied Research Laboratories (NARLabs) in Taiwan for providing computational and storage resources and the ASIAA for in-house access to high-performance computing facilities.

\end{acknowledgments}

\bibliographystyle{aasjournal}
\bibliography{main}

\end{document}